\pdfoutput=1
\documentclass[conference]{IEEEtran}
\IEEEoverridecommandlockouts
\usepackage{cite}
\usepackage{amsmath,amssymb,amsfonts}
\usepackage{graphicx}
\usepackage{textcomp}
\usepackage{xcolor}
\usepackage{url}

\usepackage{tikz}
\usetikzlibrary{shapes.geometric, arrows.meta, positioning, fit, backgrounds, calc, decorations.pathreplacing}

\newcommand{\rqanswer}[2]{\par\medskip\noindent\fbox{\parbox{\dimexpr\columnwidth-2\fboxsep-2\fboxrule\relax}{\textbf{Answer to RQ#1:} #2}}\par\medskip}

\def\BibTeX{{\rm B\kern-.05em{\sc i\kern-.025em b}\kern-.08em
		T\kern-.1667em\lower.7ex\hbox{E}\kern-.125emX}}
\begin{document}
	
	\title{InterceptFuzz: Interception-based Stateful Fuzzing of Single-Process Targets}
	
	\author{\IEEEauthorblockN{Ilja Siro\v{s}}
		\IEEEauthorblockA{
			\textit{COSIC, KU Leuven}\\
			Leuven, Belgium \\
			ilja.siros@kuleuven.be}
		\and
		\IEEEauthorblockN{Dave Singel\'ee}
		\IEEEauthorblockA{
			\textit{DistriNet, KU Leuven}\\
			Leuven, Belgium \\
			dave.singelee@kuleuven.be}
		\and
		\IEEEauthorblockN{Bart Preneel}
		\IEEEauthorblockA{
			\textit{COSIC, KU Leuven}\\
			Leuven, Belgium \\
			bart.preneel@kuleuven.be}
	}

	\maketitle
	
	\begin{abstract}
		Stateful protocol implementations are commonly fuzzed by tools that act as a network client and exchange messages with the target over sockets. Some wireless and IoT protocol stacks, in contrast, provide a single-process simulation platform, in which all nodes run within one binary and exchange packets in memory. Such simulations execute deterministically and at high speed, which makes them well suited to fuzzing, but they offer no socket to which existing stateful fuzzers can attach. At the same time, widely deployed fuzzing infrastructure, such as Google's OSS-Fuzz, supports only stateless fuzzing engines, which treat every input in isolation and offer no built-in way to reach deeper protocol states. To address both gaps, this paper presents InterceptFuzz, a middleware that enables stateful fuzzing of single-process targets relying solely on a stateless fuzzing engine, so that it can use the engines already deployed in infrastructure such as OSS-Fuzz. InterceptFuzz requires only a minimal change to the target code: a hook placed where the target passes packets between nodes. Through this hook, InterceptFuzz records the benign traffic from which the initial seeds are derived and injects the fuzzed packets. InterceptFuzz organizes the observed packet sequences in a Prefix Tree. Each seed carries a State ID that selects a node of this tree, and InterceptFuzz deterministically replays the recorded prefix before delivering the mutated packet. Over the course of a campaign, the tree grows with newly discovered sequences, while redundant states are pruned to keep its growth in check. InterceptFuzz powered by AFL++ is evaluated on the Nexus test framework of OpenThread and on the LTE and Wi-Fi models of the ns-3 network simulator, against an interception-based adaptation of AFLNet and against Concat, a baseline fuzzer also powered by AFL++ that mutates a whole message sequence as a single input and splits it into messages at a delimiter. On OpenThread and ns-3 Wi-Fi, InterceptFuzz covers significantly more branches than both baselines. During the evaluation, seven previously unknown crash signatures were found in OpenThread. The evaluation also shows that Concat, despite its simplicity, covers more branches than AFLNet in every campaign on OpenThread and ns-3 Wi-Fi, and on OpenThread it exposes the same crash signatures as InterceptFuzz, although some of them in fewer campaigns.
	\end{abstract}
	
	\begin{IEEEkeywords}
		fuzzing, stateful fuzzing, protocol testing, network simulation, wireless protocols, OpenThread, ns-3
	\end{IEEEkeywords}
	
	\section{Introduction}
	OSS-Fuzz, Google's continuous fuzzing service for open-source software, has helped identify and fix over 13,000 vulnerabilities and 50,000 bugs across 1,000 projects~\cite{OSS_Fuzz}. Yet every engine it supports---libFuzzer~\cite{libfuzzer}, AFL++~\cite{aflplusplus}, Honggfuzz~\cite{honggfuzz}, and Centipede~\cite{centipede}---is stateless. For programs that process each input independently, this is sufficient. For protocol implementations, however, whose stateful logic is reachable only through a specific sequence of prior messages, these engines offer no built-in support.
	
	This limitation reflects a broader divide in the fuzzing landscape. While stateful fuzzers exist in academia, several drawbacks hinder their industrial adoption. Existing stateful fuzzers almost universally assume a classic client-server architecture, acting as a malicious client that sends packets over OS-level network sockets. Furthermore, they are predominantly research prototypes, and many are not maintained beyond their initial publication. Adapting these tools to novel protocols or non-networked architectures therefore requires substantial manual engineering effort. A prominent class of targets highlighting this gap comprises systems in which stateful communication occurs entirely within a single process. This architectural pattern enables deterministic, high-throughput testing and is used in in-process simulators for network and IoT protocols.
	
	To bridge these gaps, this paper presents InterceptFuzz, a stateful fuzzing middleware designed for single-process environments. InterceptFuzz is not a fuzzer itself but a library linked into the target. It delegates input mutation and scheduling to an existing stateless fuzzing engine, such as AFL++ or libFuzzer. This architectural choice mitigates the maintenance issues of academic fuzzers, as InterceptFuzz inherits new features of the underlying stateless engine without requiring any modifications to it. For the same reason, InterceptFuzz can use the engines already deployed in OSS-Fuzz, a prerequisite for bringing stateful fuzzing to its targets.
	
	InterceptFuzz interacts with the target through interception hooks, i.e., single-line calls into the InterceptFuzz library that a developer places at any point of interest in the code. These hooks are designed to serve two purposes: monitoring benign traffic and injecting fuzzed inputs into the execution flow.
	From the traffic observed through these hooks, InterceptFuzz incrementally builds a Prefix Tree, a record of the packet sequences exchanged by the target. The tree grows by periodically adding the minimized corpus of the underlying stateless engine. To keep this growth in check, states that are no longer used are pruned. Each input generated by the stateless engine carries a State~ID that points to a node in the Prefix Tree. The input is delivered only after InterceptFuzz replays the recorded prefix leading to that node, so mutations never corrupt the path to the targeted state. This mechanism ensures that inputs constructed by the underlying stateless engine reach deep, sequence-dependent code paths.
	
	InterceptFuzz is evaluated on three targets built on two single-process protocol simulators: the LTE and Wi-Fi modules of the ns-3 network simulator~\cite{ns3}, which is used mainly in networking research, and the Nexus test framework of OpenThread, an open-source Thread stack used in commercial products~\cite{OpenThread}. On these targets, it is compared against two baselines. The first baseline is a delimiter-based concatenation fuzzer, referred to as Concat in the following. It implements the seed concatenation strategy described in Section~\ref{sec:stateless_bypasses} and represents the most straightforward method to perform stateful testing with a stateless engine. In this evaluation, both InterceptFuzz and Concat use AFL++ as their underlying engine. The second baseline represents dedicated stateful fuzzers and is based on AFLNet~\cite{AFLNet}, the first code- and state-coverage-guided protocol fuzzer and the reference point on which much of the subsequent work in stateful fuzzing builds~\cite{meng2025aflnet}. Specifically, the evaluation uses the LibAFL~\cite{libafl} implementation of AFLNet released with Fizzle~\cite{fizzle}, which reproduces AFLNet's message-sequence inputs and response-based state-model inference, combined with the packet-level mutations and power scheduling of LibAFL. For this evaluation, its socket communication is replaced by the same interception points used by InterceptFuzz. AFLNet distinguishes protocol states by the response codes of the target, such as the status codes of FTP or RTSP, which the protocols evaluated in this paper do not provide. Following the recommendation of the AFLNet authors~\cite{meng2025aflnet}, a dissector written for each target therefore maps every intercepted response to a code that identifies its message type. All three fuzzers share one delivery mechanism, so that the comparison isolates how each handles protocol state.

	The contributions of this paper are summarized as follows:
	\begin{itemize}
		\item InterceptFuzz, an interception-based middleware that enables stateful fuzzing of single-process targets on top of unmodified stateless fuzzing engines, such as AFL++ or libFuzzer. Unlike existing stateful fuzzers, it requires no client-server setup or network sockets: it integrates with a target through a single hook function, called where packets pass between nodes, requires only slight modifications of the target code, and can reuse the existing integration tests of the target as fuzzing harnesses.
		\item A Prefix Tree state model with a compact State~ID seed encoding, which lets a stateless engine address previously observed protocol states and reach them deterministically, together with an epoch-based loop that alternately extends, exploits, and prunes this model.
		\item An empirical evaluation on three single-process simulated targets (ns-3 LTE, ns-3 Wi-Fi, and OpenThread Nexus) against a concatenation baseline (Concat) and an interception-based adaptation of AFLNet. On OpenThread Nexus and ns-3 Wi-Fi, InterceptFuzz covers significantly more branches than both baselines.
		\item Seven previously unknown crash signatures in OpenThread.
	\end{itemize}
	
	\textbf{Responsible disclosure and open source.} All seven crash signatures found in OpenThread were reported to the OpenThread maintainers, who have since fixed them. InterceptFuzz and all experimental artifacts are released as open source at \url{https://github.com/KULeuven-COSIC/InterceptFuzz}.
	
	\section{Background \& Related Work}
	\label{sec:background}
	
	\subsection{Foundations: From Stateless to Stateful Fuzzing}
	Fuzzing, first introduced by Miller et al.~\cite{Miller_Fuzzing}, is a fundamental technique for automated vulnerability discovery. It works by feeding unexpected or malformed inputs to a target program while monitoring for abnormal behavior, such as crashes, hangs, or assertion violations. Contemporary fuzzers like AFL++ and libFuzzer employ coverage-guided mutation: rather than generating inputs from scratch, they mutate a seed corpus and track code execution. Inputs uncovering new code paths are prioritized and added to the corpus, while redundant seeds are discarded. Crucially, these conventional engines operate under a stateless assumption, treating each input as an isolated, self-contained test case.
	
	While stateless fuzzers excel at testing independent input parsers, they fall short when applied to stateful systems. In stateful software, the processing of any given input depends on the execution history and the internal state built by prior interactions. Testing such targets with isolated, stateless inputs is ineffective and inevitably leaves deeper functional logic unexplored. Consequently, stateful fuzzing must treat inputs not as single test cases, but as ordered sequences of inputs that systematically drive the target through prerequisite states before injecting mutations into those deeper execution paths.
	
	\subsection{Stateful Paradigms via Stateless Engines}
	\label{sec:stateless_bypasses}
	
	Stateless engines can be applied to stateful targets without modification if the handling of state is moved elsewhere: either into the structure of the input, or into the test harness. Both strategies are used in practice, and the second one is the closest precursor of InterceptFuzz.
	
	The \textbf{Input-Encoded Sequences (Seed Concatenation)} strategy moves state handling into the input by packing several sequential messages into a single buffer, separated by a delimiter (e.g., \texttt{message1<delim>}\dots\texttt{<delim>messageN})~\cite{daniele2024statefulfuzzingreallychallenging}.
	During execution, a custom wrapper function parses the mutated seed, splits the byte array along delimiter boundaries, and delivers the discrete packets sequentially to the target. While this approach is easy to set up, stateless engines lack semantic awareness of the delimiters and cannot focus mutations on specific packets. Consequently, standard stateless mutations frequently corrupt the delimiter tokens or invalidate early prefix messages. This fragility disrupts the message sequence and limits the fuzzer's ability to explore deeper execution states~\cite{daniele2024statefulfuzzingreallychallenging}.
	
	The \textbf{Harness-Driven State Driver} strategy moves state handling into the harness. The simplest approach is to manually write a custom wrapper (or harness) that programmatically drives the target into a specific state before delivering the fuzzed payload.
	While this protects the prerequisite trace from mutation and ensures deterministic state reachability, it scales poorly, as the analyst must manually decide whether a state is interesting and write dedicated harnesses for each state of interest. Furthermore, this strategy precludes the discovery of vulnerabilities triggered by complex relations between different mutated messages within the same trace.
	
	To give an early intuition of its design, InterceptFuzz can be viewed as an automated variant of the harness-driven strategy. Like harness-driven state drivers, it protects the prerequisite trace from mutation, avoiding the fragility of seed concatenation. Unlike manually written harnesses, however, it learns these traces automatically from observed traffic and lets them accumulate mutated messages over time. Section~\ref{sec:architecture} describes the architecture of InterceptFuzz in detail.
	
	\subsection{Stateful Fuzzing Engines}
	\label{sec:stateful_engines}
	Dedicated stateful fuzzers augment traditional stateless engines with state-aware feedback. AFLNet~\cite{AFLNet} infers protocol states from server response codes, and StateAFL~\cite{stateafl} derives them from snapshots of long-lived memory. SGFuzz~\cite{sgfuzz} instead identifies enum-typed state variables in the source code and tracks the sequences of values they take. ChatAFL~\cite{chatafl} further uses large language models to construct grammars for each message type and to mutate messages or predict the next message in a sequence. All of these approaches achieve statefulness by modifying an existing stateless fuzzing engine: AFLNet, StateAFL, and ChatAFL all derive from AFL, while SGFuzz extends libFuzzer. As a result, they do not automatically benefit from later advances in stateless fuzzing, such as those in AFL++.
	
	For standalone, network-accessible server applications, socket-based tools such as AFLNet and StateAFL offer an effective testing strategy, but they suffer from performance bottlenecks caused by OS-level network sockets, protocol timeouts, and continuous process teardowns~\cite{snapfuzz,fizzle}.
	To reduce this overhead, researchers have explored two established acceleration strategies, each with its own trade-offs, and a more recent approach has proposed a third.
	
	The first strategy removes the kernel network stack from the fuzzing loop by intercepting the target's system or library calls in user space. SnapFuzz~\cite{snapfuzz} uses load-time binary rewriting to replace Internet sockets with UNIX domain sockets and make network communication synchronous, snapshots the target at the latest safe point, and redirects file operations to an in-memory filesystem, reporting speedups of up to 62.8$\times$ over AFLNet. Lightweight preload libraries such as \texttt{libdesock}~\cite{libdesock} follow the same principle, replacing socket APIs with in-memory emulation. Fizzle~\cite{fizzle} goes further, preloading a shared library that emulates sockets, threading, timing, and entropy sources in user space, serializing multithreaded servers to obtain fully deterministic and reproducible executions at substantially higher speed. Such executors, however, address only how inputs are delivered: stateful input generation is left to the fuzzing engine, and inputs are still exchanged through (emulated) socket endpoints of a client-server application.
	
	The second strategy relies on hypervisor-based systems, such as Nyx-Net~\cite{nyxnet}, which utilize rapid virtual machine snapshotting to avoid standard network I/O. While this resolves the reset bottleneck, it imposes heavy deployment overheads, requiring custom kernels, modified hypervisors, and hardware-assisted virtualization.
	
	A third, more recent approach, LibAFLstar~\cite{LibAFLStar}, keeps the server running in persistent mode, so that many messages can be fuzzed in a selected state without restarts. As its authors note, however, a fuzzed message may move the server into another state, so later messages are not guaranteed to reach the selected state.
	
	All of these approaches, however, remain tied to the socket-based client-server model. Section~\ref{sec:bg:simulation} discusses a class of targets for which this assumption does not hold.
	
	\subsection{Interception-Based Fuzzing}
	\label{sec:bg:interception}
	
	The stateful engines discussed above act as an external peer that delivers each message as an opaque byte buffer through the target's network interface, typically a socket. This abstraction becomes a liability for multi-layer protocol stacks, which are the norm in wireless and IoT communication. To fuzz a single layer while leaving the rest of the stack intact, every mutated message must still carry valid lower-layer framing, sequence numbers, checksums and, where applicable, encryption. Achieving this from outside the target requires custom mutators or wrappers that re-encapsulate each mutated payload. Such components are protocol-specific and must be rewritten for every new target.
	
	Interception-based fuzzing avoids this problem by placing the fuzzer inside the communication path. Rather than generating complete messages externally, it installs a hook at a point where the target produces or consumes packets and operates on packets in flight: the protocol implementation itself generates benign traffic, and the hook selectively modifies, drops, or replaces packets at the chosen layer before they continue along the normal processing path. Three properties of this design are relevant to this work. First, mutations are \emph{layer-selective}: since the hook sits at a specific layer boundary, the surrounding stack performs encapsulation, and the fuzzer never has to reconstruct well-formed outer headers. Second, the approach is \emph{transport-agnostic}: the hook can be placed on a radio link, at a driver boundary, or inside a single process, without assuming OS-level sockets. Third, a single hook serves both \emph{observation and injection}: it can record unmodified traffic to capture how the protocol behaves, and later overwrite packets with fuzzed payloads.
	
	Interception is widely used in wireless protocol fuzzing, with tools targeting Wi-Fi (Greyhound~\cite{greyhound}), Bluetooth Low Energy (SweynTooth~\cite{sweyntooth}, BLuEMan~\cite{BLuEMan}), Bluetooth Classic (BrakTooth~\cite{braktooth}), Bluetooth controller firmware (Frankenstein~\cite{Frankenstein}), 5G NR (5Ghoul~\cite{5ghoul}), 4G LTE and 5G NR (CovFUZZ~\cite{covfuzz}), LTE-M and NB-IoT~\cite{breakingsilence}, and Thread (ThreadFuzzer~\cite{11659538}). Beyond wireless protocols, interception has also been applied at the hardware--OS boundary of device drivers (PeriScope~\cite{PeriScope}).

	Despite this breadth, existing interception-based protocol fuzzers share two limitations. First, each is built for a single protocol, such as Wi-Fi, Bluetooth, 5G NR, or Thread, and applying it to a different protocol requires additional protocol-specific engineering. Second, to the best of the authors' knowledge, none of them maintains an explicit model of observed states in which a particular state can be selected and reached deterministically by replaying recorded traffic. Instead, deep states are reached through live interaction with a peer device, physical or simulated, sometimes guided by a protocol-specific state model that is either written manually from the specification (Greyhound) or learned from traces using hand-crafted mapping rules (5Ghoul). The following subsection examines a class of targets in which both limitations matter most.
	
	\subsection{In-Process Protocol Simulation}
	\label{sec:bg:simulation}
	
	Wireless and IoT protocol stacks such as 4G/5G, Wi-Fi, Bluetooth, and Thread are ultimately deployed on dedicated radio hardware, yet testing them on physical devices is expensive, slow, and difficult to reproduce: testbeds are costly to scale, over-the-air links introduce nondeterministic loss and timing, and the internal state of a device is hard to observe. For this reason, mature protocol projects commonly provide a simulation platform that replaces the radio and hardware drivers with a software model, allowing the protocol logic to be exercised on a commodity machine.
	Such platforms follow one of two architectures. 
	
	In \emph{multi-process} simulations, each simulated node runs as a separate OS process, and nodes exchange frames over an inter-process channel. OpenThread's POSIX simulation platform follows this model, running one process per device with the IEEE~802.15.4 radio driver implemented on top of UDP sockets~\cite{OpenThread_Simulation}. Open-source cellular stacks such as srsRAN similarly connect separate user-equipment and base-station processes through a virtual radio link~\cite{srsRAN_ZMQ}, and the same holds for the 5G stack of the srsRAN Project~\cite{srsRAN_Project_ZMQ} and for the Open5GS core network, whose test suite starts every network function as a separate process~\cite{open5gs_tests}. For Bluetooth, Zephyr is simulated with BabbleSim, which runs every device and the physical layer in separate processes~\cite{zephyr_bsim}, and the Wi-Fi test framework of hostapd and wpa\_supplicant combines the \texttt{mac80211\_hwsim} kernel module with several user-space processes~\cite{hostap_hwsim}. From a fuzzing perspective, these setups inherit the drawbacks discussed in Section~\ref{sec:stateful_engines}: every execution involves inter-process communication, multiple process restarts, and timing coordination across processes.
	
	In \emph{single-process} simulations, by contrast, all nodes and the medium connecting them reside in one address space, frames are passed between node models as in-memory objects, and execution advances along a discrete virtual timeline rather than wall-clock time. Resetting the entire simulated network therefore amounts to resetting a single process, one binary exposes the code of every node to coverage instrumentation, and, given a fixed random seed, execution is fully deterministic. These properties make single-process simulations particularly well suited to high-throughput fuzzing, yet they are the least accessible to existing tools. Socket-based stateful fuzzers and their accelerated variants have no socket to attach to, and the interception approaches reviewed in Section~\ref{sec:bg:interception} are each built for a single protocol. Moreover, even where an interception point can be placed, these approaches provide no means to select a particular state and reach it deterministically. What is missing is a generic interception mechanism that can be placed at an arbitrary in-memory packet boundary and that serves not only to mutate traffic but also to route the target deterministically into previously observed states. InterceptFuzz, described in the next section, is designed to fill this gap.
	
	Among open-source wireless protocol stacks, at the time of research, single-process simulators remain rare. This paper evaluates InterceptFuzz on two such targets. ns-3~\cite{ns3} is a discrete-event network simulator widely used in networking research. It executes complete topologies---including link-layer models, routing protocols, and transport protocols---inside a single process, with packets exchanged between simulated nodes through channel objects and scheduled on an internal event queue. OpenThread~\cite{OpenThread} is an open-source implementation of the Thread mesh networking protocol. In addition to its multi-process POSIX simulation, OpenThread provides the Nexus test framework, in which multiple Thread nodes run within a single process and exchange IEEE~802.15.4 frames through in-memory buffers. This environment is used in the evaluation.

	\section{InterceptFuzz Architecture}
	\label{sec:architecture}
	\subsection{High-Level Overview}
	\label{sec:overview}
	InterceptFuzz is a middleware designed to enable stateful fuzzing by leveraging standard stateless engines such as AFL++ or libFuzzer. Rather than operating as an independent fuzzer, it is compiled directly into the target binary as a linked library. This design is tailored specifically for targets exhibiting in-process stateful communication, such as mock-I/O integration tests and single-process network simulators. InterceptFuzz assumes that such targets execute deterministically for a given input, i.e., replaying a recorded prefix triggers the same sequence of hook invocations and drives the target into the same state. Single-process simulators such as ns-3 and the Nexus framework of OpenThread satisfy this assumption when run with a fixed random seed, as discussed in Section~\ref{sec:bg:simulation}. Determinism also allows recorded packets containing freshness-dependent fields, such as frame counters or message integrity codes, to be replayed verbatim.
	
	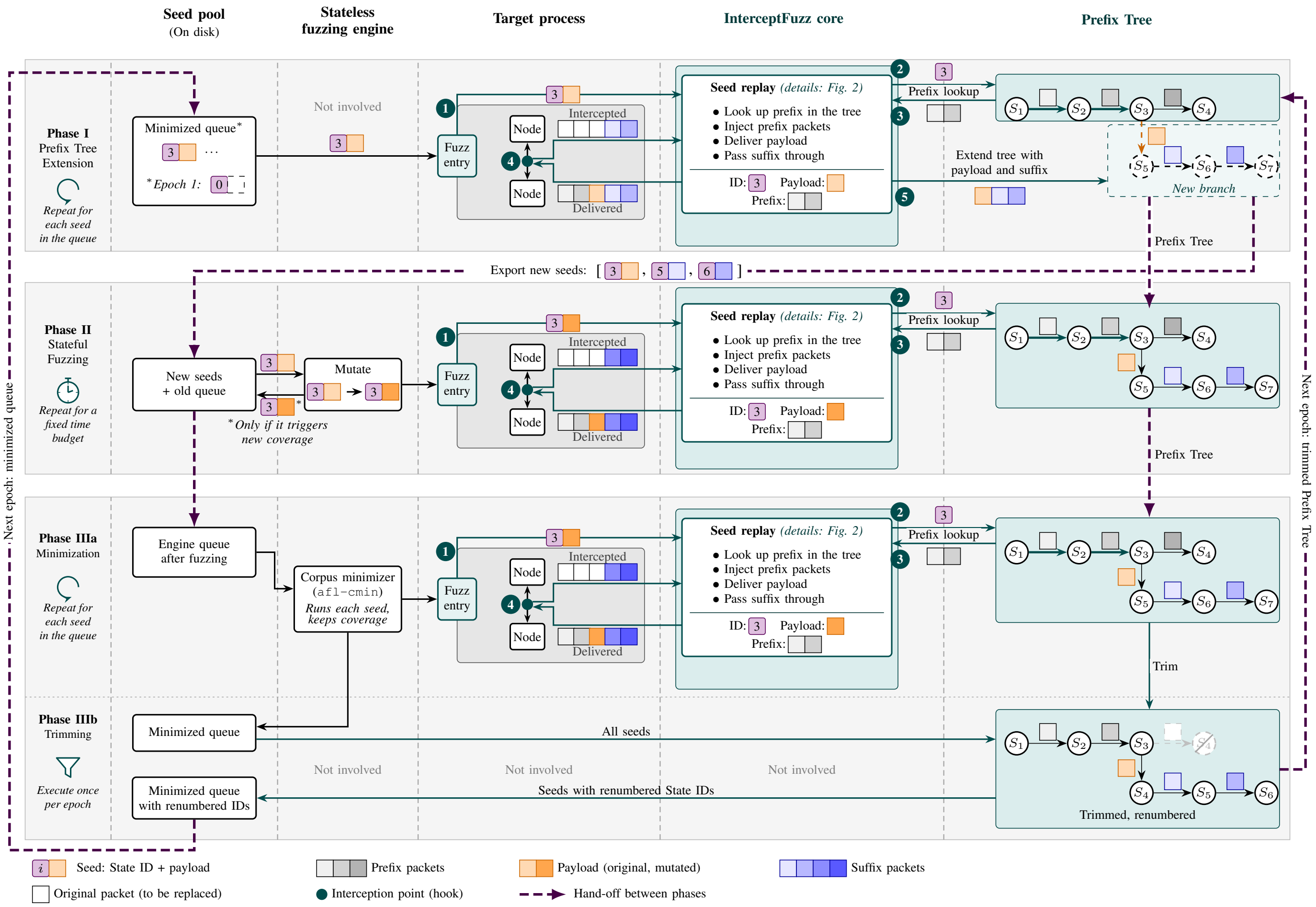
\begin{figure*}[t]
		\centering
\newcommand{\seedreplay}[2]{%
 \draw[cbox] ({\xB+0.42},-1.0) rectangle ({\xP-0.3},1.65);
 \node[font=\scriptsize] at ({(\xB+\xP)/2+0.06},1.4) {\textbf{Seed replay} {\color{teal!45!black}\itshape(details: Fig.~\ref{fig:running-example})}};
 \node[font=\scriptsize,align=left,anchor=north] at ({(\xB+\xP)/2+0.06},1.17) {$\bullet$ Look up prefix in the tree\\$\bullet$ Inject prefix packets\\$\bullet$ Deliver payload\\$\bullet$ Pass suffix through};
 \draw[teal!40!black,thin] ({\xB+0.57},-0.16) -- ({\xP-0.45},-0.16);
 \node[font=\scriptsize,inner sep=0pt] at ({(\xB+\xP)/2+0.06},-0.42) {ID:\kern1pt\tikz[baseline=-0.6ex]\node[idb]{#1};\hspace{8pt}Payload:\kern1pt\tikz[baseline=-0.6ex]\node[#2]{};};
 \node[font=\scriptsize,inner sep=0pt] at ({(\xB+\xP)/2+0.06},-0.77) {Prefix:\kern1pt\tikz[baseline=-0.6ex]{\node[cA] (pa) {}; \node[cB,anchor=west] at ([xshift=-0.4pt]pa.east) {};}};}
\newcommand{\tgt}[5]{%
 \node[nd] (#2n1) at ({\xT+2.1},0.62) {Node};
 \node[nd] (#2n2) at ({\xT+2.1},-0.62) {Node};
 \node[hook] (#2hk) at ({\xT+2.1},0) {};
 \draw[{Stealth[length=1.6mm]}-{Stealth[length=1.6mm]},thick] (#2n1) -- (#2n2); \node[hook] at (#2hk) {};
 \node[num] at ($(#2hk)+(-0.32,0)$) {4};
 \draw[ea] ($(#2hk)+(0.1,0.05)$) -- ++(0.4,0) |- ({\xB+0.42},0.4);
 \draw[ea] ({\xB+0.42},-0.4) -| ($(#2hk)+(0.5,-0.05)$) -- ($(#2hk)+(0.1,-0.05)$);
 \foreach \st/\k in {wh/0,wh/1,wh/2,#3/3,#4/4} \node[\st] at ({\xT+2.85+0.3*\k},0.62) {};
 \foreach \st/\k in {cA/0,cB/1,#5/2,#3/3,#4/4} \node[\st] at ({\xT+2.85+0.3*\k},-0.62) {};
 \node[font=\scriptsize,text=black!75] at ({\xT+3.45},0.9) {Intercepted};
 \node[font=\scriptsize,text=black!75] at ({\xT+3.45},-0.9) {Delivered};}
		\resizebox{\textwidth}{!}{%
\begin{tikzpicture}[
  font=\footnotesize,
  >={Stealth[length=2mm]},
  ext/.style={draw,thick,rounded corners=2pt,align=center,fill=white,font=\scriptsize},
  ifz/.style={draw=teal!60!black,thick,rounded corners=2pt,align=center,fill=teal!10,font=\scriptsize},
  nd/.style={draw,thick,rounded corners=1.5pt,fill=white,inner xsep=1.5pt,minimum width=6.6mm,minimum height=5mm,font=\scriptsize},
  hook/.style={circle,draw=teal!60!black,fill=teal!60!black,minimum size=2mm,inner sep=0pt},
  pk/.style={rectangle,draw,minimum width=2.4mm,minimum height=2.4mm,inner sep=0pt},
  pre/.style={pk,fill=gray!25},
  pay/.style={pk,fill=orange!35,draw=orange!80!black},
  suf/.style={pk,fill=blue!15,draw=blue!60!black},
  org/.style={pk,fill=white},
  st/.style={circle,draw,thick,minimum size=4.4mm,inner sep=0pt,font=\scriptsize,fill=white},
  e/.style={->,thick},
  ea/.style={->,thick,teal!60!black},
  lab/.style={font=\scriptsize,inner sep=1.5pt,align=center},
  num/.style={circle,fill=teal!60!black,text=white,font=\scriptsize\bfseries,inner sep=0pt,minimum size=3.7mm},
  cbox/.style={draw=teal!60!black,thick,rounded corners=2pt,fill=white},
  idb/.style={rectangle,rounded corners=1.2pt,draw=violet!70!black,fill=violet!25,minimum width=3.2mm,minimum height=3.2mm,inner sep=0pt,font=\scriptsize},
  lp/.style={minimum width=3.2mm,minimum height=3.2mm,font=\tiny},
  pics/seed/.style n args={2}{code={\node[idb] (-i) at (0,0) {#1}; \node[pay,lp,anchor=west] at ([xshift=-0.4pt]-i.east) {#2};}},
  el/.style={pk,lp,fill=gray!25,draw=black,thin,solid,text=black},
  elp/.style={pk,lp,fill=orange!35,draw=orange!80!black,thin,solid,text=black},
  elo/.style={pk,lp,fill=white,draw=black,thin,solid,text=black},
  cA/.style={pk,lp,thin,solid,draw=black,fill=gray!12},
  cB/.style={pk,lp,thin,solid,draw=black,fill=gray!35},
  cC/.style={pk,lp,thin,solid,draw=black,fill=gray!60},
  cD/.style={pk,lp,thin,solid,draw=orange!80!black,fill=orange!30},
  cDp/.style={pk,lp,thin,solid,draw=orange!80!black,fill=orange!70},
  cE/.style={pk,lp,thin,solid,draw=blue!60!black,fill=blue!10},
  cF/.style={pk,lp,thin,solid,draw=blue!60!black,fill=blue!28},
  cG/.style={pk,lp,thin,solid,draw=blue!60!black,fill=blue!45},
  cH/.style={pk,lp,thin,solid,draw=blue!60!black,fill=blue!65},
  pics/seedc/.style n args={2}{code={\node[idb] (-i) at (0,0) {#1}; \node[#2,anchor=west] at ([xshift=-0.4pt]-i.east) {};}},
  hand/.style={-{Latex[length=2.8mm,width=2.2mm]},line width=1.4pt,violet!55!black,dash pattern=on 5pt off 2.5pt},
  wh0/.style={pk,lp,thin,draw=black,dashed,fill=white},
  wh/.style={pk,lp,thin,draw=black,fill=white},
  tgt/.style={st,draw=violet!70!black,fill=violet!25,very thick},
  sub/.style={rounded corners=1.5pt,fill=gray!60!black,text=white,font=\tiny\bfseries,inner sep=0pt,minimum size=3.2mm},
  rule/.style={font=\scriptsize,anchor=west,inner sep=1pt},
]
\def\xL{0}\def\xS{1.65}\def\xE{4.85}\def\xT{7.55}\def\xB{12.2}\def\xP{16.95}\def\xR{24.3}
\def\hdr{12.73}
\foreach \yb/\yt in {-3.1/3.48,3.92/7.6,8.2/11.88} {\fill[gray!8] (\xL,\yb) rectangle (\xR,\yt); \draw[gray!55] (\xL,\yb) rectangle (\xR,\yt);
  \foreach \x in {\xS,\xE,\xT,\xB,\xP+0.7} \draw[gray!70,semithick,dash pattern=on 3pt off 2pt] (\x,\yb) -- (\x,\yt);}
\draw[gray!70,thin,dash pattern=on 1pt off 1.5pt] (\xL,-0.36) -- (\xR,-0.36);
\foreach \yb/\yt in {4.02/7.5,8.3/11.76} {
  \filldraw[fill=teal!14,draw=teal!55!black,rounded corners=2pt] ({\xB+0.3},\yb) rectangle ({\xP-0.18},\yt);
}
\foreach \yb/\yt in {1.07/3.08,-2.88/-0.6,5.19/7.2,10.7/11.6} \filldraw[fill=teal!14,draw=teal!55!black,rounded corners=2pt] ({\xP+1.7},\yb) rectangle ({\xR-0.2},\yt);
\filldraw[fill=teal!6,draw=teal!55!black,dashed,rounded corners=2pt] ({\xP+3.85},9.25) rectangle ({\xR-0.2},10.63);
\node[lab,font=\footnotesize\bfseries] at ({(\xS+\xE)/2},\hdr) {Seed pool};
\node[lab] at ({(\xS+\xE)/2},\hdr-0.33) {(On disk)};
\node[lab,font=\footnotesize\bfseries] at ({(\xE+\xT)/2},\hdr-0.12) {Stateless\\fuzzing engine};
\node[lab,font=\footnotesize\bfseries] at ({(\xT+\xB)/2},\hdr-0.08) {Target process};
\node[lab,font=\footnotesize\bfseries,text=teal!45!black] at ({(\xB+\xP)/2},\hdr-0.08) {InterceptFuzz core};
\node[lab,font=\footnotesize\bfseries,text=teal!45!black] at ({(\xP+0.7+\xR)/2},\hdr-0.08) {Prefix Tree};
\node[lab] at ({(\xL+\xS)/2},10.18) {\textbf{Phase I}\\Prefix Tree\\Extension};
\begin{scope}[shift={({(\xL+\xS)/2},9.38)}]
 \draw[->,thick,teal!60!black,>={Stealth[length=1.6mm]}] (0.2,0) arc (0:290:0.2);
\end{scope}
\node[lab,font=\fontsize{6.4}{7.6}\selectfont\itshape] at ({(\xL+\xS)/2},8.7) {Repeat for\\each seed\\in the queue};
\node[lab] at ({(\xL+\xS)/2},6.39) {\textbf{Phase II}\\Stateful\\Fuzzing};
\begin{scope}[shift={({(\xL+\xS)/2},5.49)}]
 \draw[thick,teal!60!black,fill=white] (0,0) circle (0.2);
 \foreach \a in {0,90,180,270} \draw[teal!60!black] (\a:0.15) -- (\a:0.19);
 \draw[thick,teal!60!black] (0,0) -- (90:0.13); \draw[thick,teal!60!black] (0,0) -- (20:0.1);
 \draw[thick,teal!60!black] (0,0.2) -- (0,0.26); \draw[thick,teal!60!black] (-0.06,0.27) -- (0.06,0.27);
\end{scope}
\node[lab,font=\fontsize{6.4}{7.6}\selectfont\itshape] at ({(\xL+\xS)/2},4.86) {Repeat for a\\fixed time\\budget};
\node[lab] at ({(\xL+\xS)/2},2.55) {\textbf{Phase IIIa}\\{\fontsize{6.4}{7.6}\selectfont Minimization}};
\begin{scope}[shift={({(\xL+\xS)/2},1.75)}]
 \draw[->,thick,teal!60!black,>={Stealth[length=1.6mm]}] (0.2,0) arc (0:290:0.2);
\end{scope}
\node[lab,font=\fontsize{6.4}{7.6}\selectfont\itshape] at ({(\xL+\xS)/2},1.07) {Repeat for\\each seed\\in the queue};
\node[lab] at ({(\xL+\xS)/2},-0.95) {\textbf{Phase IIIb}\\{\fontsize{6.4}{7.6}\selectfont Trimming}};
\begin{scope}[shift={({(\xL+\xS)/2},-1.7)}]
 \draw[thick,teal!60!black,fill=white,line join=round] (-0.22,0.17) -- (0.22,0.17) -- (0.05,-0.03) -- (0.05,-0.2) -- (-0.05,-0.14) -- (-0.05,-0.03) -- cycle;
\end{scope}
\node[lab,font=\fontsize{6.4}{7.6}\selectfont\itshape] at ({(\xL+\xS)/2},-2.3) {Execute once\\per epoch};

\begin{scope}[yshift=9.93cm]
 \filldraw[fill=gray!20,draw=gray!65!black,rounded corners=2pt] ({\xT+0.75},-1.12) rectangle ({\xB-0.3},1.08);
 \node[ext,minimum width=23.6mm,minimum height=17mm,inner sep=0pt] (pool1) at ({(\xS+\xE)/2},0.0) {};
 \node[font=\scriptsize] at ($(pool1.north)+(0,-0.24)$) {Minimized queue$^{*}$};
 \node[font=\scriptsize\itshape,anchor=east] at ($(pool1.center)+(0.26,-0.45)$) {$^{*}$Epoch~1:};
 \pic at ($(pool1.center)+(0.48,-0.45)$) {seedc={0}{wh0}};
 \node[ifz,minimum height=8mm] (fe1) at ({\xT+0.75},0.1) {Fuzz\\entry};
 \tgt{fe1}{a}{cE}{cF}{cD}
 \seedreplay{3}{cD}
 \draw[ea] (fe1.north) |- ({\xB+0.42},1.28);
 \pic at ({(\xT+\xB)/2+0.3},1.28) {seedc={3}{cD}};
 \node[num] at ($(fe1.north)+(-0.22,0.5)$) {1};
 \draw[ea] ({\xP-0.3},1.47) -- node[num,above=3pt,pos=0.08]{2} ({\xP+1.7},1.47);
 \node[idb,anchor=south] at ({\xP+0.7},1.55) {3};
 \node[font=\scriptsize,align=center,fill=gray!8,inner sep=1pt] at ({\xP+0.7},1.32) {Prefix lookup};
 \draw[ea] ({\xP+1.7},1.17) -- node[num,below=3pt,pos=0.92]{3} ({\xP-0.3},1.17);
 \node[cA,anchor=north] at ({\xP+0.54},1.09) {}; \node[cB,anchor=north] at ({\xP+0.86},1.09) {};
 \draw[e] (pool1.east |- fe1.west) -- node[pos=0.5,above=2pt,inner sep=0pt,minimum width=6.4mm,minimum height=3.2mm] (s1c) {} (fe1.west);
 \pic at ($(s1c.center)+(-0.16,0)$) {seedc={3}{cD}};
 \node[lab,text=gray] at ({(\xE+\xT)/2},1.05) {Not involved};
 \pic at ($(pool1.center)+(-0.45,0.17)$) {seedc={3}{cD}}; \node[font=\scriptsize] at ($(pool1.center)+(0.35,0.17)$) {\dots};
 \begin{scope}[shift={(\xP+2.1,1.0)}]
  \node[st] (t0) at (0,0) {$S_1$}; \node[st] (t1) at (1.2,0) {$S_2$}; \node[st] (t2) at (2.4,0) {$S_3$}; \node[st] (t3) at (3.6,0) {$S_4$};
  \node[st,dashed] (t4) at (2.4,-1.1) {$S_5$}; \node[st,dashed] (t5) at (3.6,-1.1) {$S_6$}; \node[st,dashed] (t6) at (4.8,-1.1) {$S_7$};
  \draw[->,very thick,teal!60!black] (t0)--(t1) node[cA,midway,above=1.5pt] {}; \draw[->,very thick,teal!60!black] (t1)--(t2) node[cB,midway,above=1.5pt] {}; \draw[->,thin] (t2)--(t3) node[cC,midway,above=1.5pt] {};
  \draw[->,thick,dashed,orange!80!black] (t2)--(t4) node[cD,solid,pos=0.45,right=3.5pt] {}; \draw[->,thick,dashed] (t4)--(t5) node[cE,solid,midway,above=1.5pt] {}; \draw[->,thick,dashed] (t5)--(t6) node[cF,solid,midway,above=1.5pt] {};
  \node[font=\scriptsize\itshape,text=teal!45!black] at (3.6,-1.52) {New branch};
 \end{scope}
 \draw[ea] ({\xP-0.3},-0.38) -- node[num,below=3pt,pos=0.06]{5} node[font=\scriptsize,above=1pt,pos=0.5,text=black,fill=gray!8,inner sep=1pt,align=center]{Extend tree with\\payload and suffix} ({\xP+3.85},-0.38);
 \node[cD,anchor=north] at ({\xP+1.46},-0.5) {}; \node[cE,anchor=north] at ({\xP+1.78},-0.5) {}; \node[cF,anchor=north] at ({\xP+2.1},-0.5) {};
\end{scope}
\draw[hand] ({\xR-0.7},9.25) |- ({(\xS+\xE)/2},7.82) -- ({(\xS+\xE)/2},6.12);

\fill[white] (8.45,7.66) rectangle (13.88,7.99);
\node[lab,anchor=east] at (10.85,7.82) {Export new seeds:};
\node[font=\small] at (11.02,7.82) {[}; \node[font=\small] at (11.89,7.76) {,}; \node[font=\small] at (12.79,7.76) {,}; \node[font=\small] at (13.73,7.82) {]};
\pic at (11.30,7.82) {seedc={3}{cD}}; \pic at (12.20,7.82) {seedc={5}{cE}}; \pic at (13.10,7.82) {seedc={6}{cF}};
\begin{scope}[yshift=5.54cm]
 \filldraw[fill=gray!20,draw=gray!65!black,rounded corners=2pt] ({\xT+0.75},-1.12) rectangle ({\xB-0.3},1.08);
 \node[ext,minimum width=23.6mm,minimum height=10mm] (pool2) at ({(\xS+\xE)/2},0.1) {New seeds\\+ old queue};
 \node[ext,minimum width=18.6mm,minimum height=10mm] (eng2) at ({(\xE+\xT)/2+0.1},0.1) {Mutate\\[9pt]};
 \pic at ($(eng2.center)+(-0.72,-0.14)$) {seedc={3}{cD}}; \draw[->,thick] ($(eng2.center)+(-0.12,-0.14)$) -- ($(eng2.center)+(0.18,-0.14)$); \pic at ($(eng2.center)+(0.4,-0.14)$) {seedc={3}{cDp}};
 \draw[e] ($(pool2.east)+(0,0.2)$) -- ($(eng2.west)+(0,0.2)$);
 \pic at ({\xE-0.16},0.52) {seedc={3}{cD}};
 \pic at ({\xE-0.16},-0.33) {seedc={3}{cDp}}; \node[font=\scriptsize,inner sep=0pt] at ({\xE+0.42},-0.29) {$^{*}$};
 \node[font=\scriptsize\itshape,align=center,fill=gray!8,inner sep=1pt] at ({\xE},-0.82) {$^{*}$Only if it triggers\\new coverage};
 \draw[e] ($(eng2.west)+(0,-0.2)$) -- ($(pool2.east)+(0,-0.2)$);
 \node[ifz,minimum height=8mm] (fe) at ({\xT+0.75},0.1) {Fuzz\\entry};
 \tgt{fe}{b}{cG}{cH}{cDp}
 \draw[e] (eng2) -- (fe);
 \seedreplay{3}{cDp}
 \draw[ea] (fe.north) |- ({\xB+0.42},1.28);
 \pic at ({(\xT+\xB)/2+0.3},1.28) {seedc={3}{cDp}};
 \node[num] at ($(fe.north)+(-0.22,0.5)$) {1};
 \draw[ea] ({\xP-0.3},1.47) -- node[num,above=3pt,pos=0.08]{2} ({\xP+1.7},1.47);
 \node[idb,anchor=south] at ({\xP+0.7},1.55) {3};
 \node[font=\scriptsize,align=center,fill=gray!8,inner sep=1pt] at ({\xP+0.7},1.32) {Prefix lookup};
 \draw[ea] ({\xP+1.7},1.17) -- node[num,below=3pt,pos=0.92]{3} ({\xP-0.3},1.17);
 \node[cA,anchor=north] at ({\xP+0.54},1.09) {}; \node[cB,anchor=north] at ({\xP+0.86},1.09) {};
 \begin{scope}[shift={(\xP+2.1,1.0)}]
  \node[st] (u0) at (0,0) {$S_1$}; \node[st] (u1) at (1.2,0) {$S_2$}; \node[st] (u2) at (2.4,0) {$S_3$}; \node[st] (u3) at (3.6,0) {$S_4$};
  \node[st] (u4) at (2.4,-0.95) {$S_5$}; \node[st] (u5) at (3.6,-0.95) {$S_6$}; \node[st] (u6) at (4.8,-0.95) {$S_7$};
  \draw[->,very thick,teal!60!black] (u0)--(u1) node[cA,midway,above=1.5pt] {}; \draw[->,very thick,teal!60!black] (u1)--(u2) node[cB,midway,above=1.5pt] {}; \draw[->,thin] (u2)--(u3) node[cC,midway,above=1.5pt] {};
  \draw[->,thin] (u2)--(u4) node[cD,midway,left=3.5pt] {}; \draw[->,thin] (u4)--(u5) node[cE,midway,above=1.5pt] {}; \draw[->,thin] (u5)--(u6) node[cF,midway,above=1.5pt] {};
  \end{scope}
\end{scope}
\draw[hand] (pool2.south) -- ({(\xS+\xE)/2},2.89);

\begin{scope}[yshift=1.42cm]
 \filldraw[fill=gray!20,draw=gray!65!black,rounded corners=2pt] ({\xT+0.75},-1.12) rectangle ({\xB-0.3},1.08);
 \filldraw[fill=teal!14,draw=teal!55!black,rounded corners=2pt] ({\xB+0.3},-1.63) rectangle ({\xP-0.18},1.83);
 \node[ext,minimum width=23.6mm,minimum height=9.5mm] (engq) at ({(\xS+\xE)/2},1.0) {Engine queue\\{\scriptsize after fuzzing}};
 \node[ext,minimum width=20.6mm,minimum height=10mm,font=\scriptsize,inner xsep=2pt] (eng3) at ({(\xE+\xT)/2},0.1) {Corpus minimizer\\(\texttt{afl-cmin})\\[1pt]{\scriptsize\itshape Runs each seed,}\\[-2pt]{\scriptsize\itshape keeps coverage}};
 \draw[e] (engq.east) -- ++(0.35,0) |- ($(eng3.west)+(0,0.2)$);
 \node[ifz,minimum height=8mm] (fe3) at ({\xT+0.75},0.1) {Fuzz\\entry};
 \tgt{fe3}{c}{cG}{cH}{cDp}
 \draw[e] (eng3.east |- fe3.west) -- (fe3.west);
 \seedreplay{3}{cDp}
 \draw[ea] (fe3.north) |- ({\xB+0.42},1.28);
 \pic at ({(\xT+\xB)/2+0.3},1.28) {seedc={3}{cDp}};
 \node[num] at ($(fe3.north)+(-0.22,0.5)$) {1};
 \draw[ea] ({\xP-0.3},1.47) -- node[num,above=3pt,pos=0.08]{2} ({\xP+1.7},1.47);
 \node[idb,anchor=south] at ({\xP+0.7},1.55) {3};
 \node[font=\scriptsize,align=center,fill=gray!8,inner sep=1pt] at ({\xP+0.7},1.32) {Prefix lookup};
 \draw[ea] ({\xP+1.7},1.17) -- node[num,below=3pt,pos=0.92]{3} ({\xP-0.3},1.17);
 \node[cA,anchor=north] at ({\xP+0.54},1.09) {}; \node[cB,anchor=north] at ({\xP+0.86},1.09) {};
 \begin{scope}[shift={(\xP+2.1,1.0)}]
  \node[st] (w0) at (0,0) {$S_1$}; \node[st] (w1) at (1.2,0) {$S_2$}; \node[st] (w2) at (2.4,0) {$S_3$}; \node[st] (w3) at (3.6,0) {$S_4$};
  \node[st] (w4) at (2.4,-0.95) {$S_5$}; \node[st] (w5) at (3.6,-0.95) {$S_6$}; \node[st] (w6) at (4.8,-0.95) {$S_7$};
  \draw[->,very thick,teal!60!black] (w0)--(w1) node[cA,midway,above=1.5pt] {}; \draw[->,very thick,teal!60!black] (w1)--(w2) node[cB,midway,above=1.5pt] {}; \draw[->,thin] (w2)--(w3) node[cC,midway,above=1.5pt] {};
  \draw[->,thin] (w2)--(w4) node[cD,midway,left=3.5pt] {}; \draw[->,thin] (w4)--(w5) node[cE,midway,above=1.5pt] {}; \draw[->,thin] (w5)--(w6) node[cF,midway,above=1.5pt] {};
 \end{scope}
\end{scope}
\node[ext,minimum width=23.6mm,minimum height=7mm] (mq) at ({(\xS+\xE)/2},-1.05) {Minimized queue};
\node[ext,minimum width=23.6mm,minimum height=8mm] (pool3) at ({(\xS+\xE)/2},-2.3) {Minimized queue\\{\scriptsize with renumbered IDs}};
\draw[e] (eng3.south) |- ($(mq.east)+(0,0.12)$);
\draw[ea] ($(mq.east)+(0,-0.12)$) -- node[lab,above,text=black,pos=0.5]{All seeds} ({\xP+1.7},-1.17);
\draw[ea] ({\xP+1.7},-2.3) -- node[lab,above,text=black,pos=0.5]{Seeds with renumbered State IDs} (pool3.east |- {(0,-2.3)});
\node[lab,text=gray] at ({(\xE+\xT)/2},-1.75) {Not involved};
\node[lab,text=gray] at ({(\xT+\xB)/2},-1.75) {Not involved};
\node[lab,text=gray] at ({(\xB+\xP+0.7)/2},-1.75) {Not involved};
\begin{scope}[shift={(\xP+2.1,-1.25)}]
  \node[st] (v0) at (0,0) {$S_1$}; \node[st] (v1) at (1.2,0) {$S_2$}; \node[st] (v2) at (2.4,0) {$S_3$};
  \node[st,draw=gray!50,text=gray!60,dashed] (v3) at (3.6,0) {$S_4$};
  \node[st] (v4) at (2.4,-0.95) {$S_4$}; \node[st] (v5) at (3.6,-0.95) {$S_5$}; \node[st] (v6) at (4.8,-0.95) {$S_6$};
  \draw[->,thin] (v0)--(v1) node[cA,midway,above=1.5pt] {}; \draw[->,thin] (v1)--(v2) node[cB,midway,above=1.5pt] {}; \draw[->,thin,gray!50,dashed] (v2)--(v3) node[pk,lp,midway,above=1.5pt,draw=gray!50,text=gray!60,fill=white,dashed] {};
  \draw[->,thin] (v2)--(v4) node[cD,midway,left=3.5pt] {}; \draw[->,thin] (v4)--(v5) node[cE,midway,above=1.5pt] {}; \draw[->,thin] (v5)--(v6) node[cF,midway,above=1.5pt] {};
  \draw[gray!70,thick] ($(v3)+(-0.17,-0.17)$) -- ($(v3)+(0.17,0.17)$);
\end{scope}
\node[lab] at ({(\xP+1.7+\xR-0.2)/2},-2.63) {Trimmed, renumbered};
\draw[ea] ({\xP+4.65},1.07) -- node[lab,right,text=black,pos=0.5,fill=gray!8]{Trim} ({\xP+4.65},-0.6);
\begin{scope}[shift={(0.3,-3.65)}]
 \pic at (0,0) {seedc={$i$}{cD}}; \node[rule] at (0.65,0) {Seed: State~ID + payload};
 \node[cA] at (5.46,0) {}; \node[cB] at (5.78,0) {}; \node[cC] at (6.1,0) {}; \node[rule] at (6.31,0) {Prefix packets};
 \node[cD] at (9.36,0) {}; \node[cDp] at (9.68,0) {}; \node[rule] at (9.89,0) {Payload (original, mutated)};
 \node[cE] at (14.36,0) {}; \node[cF] at (14.68,0) {}; \node[cG] at (15.0,0) {}; \node[cH] at (15.32,0) {}; \node[rule] at (15.53,0) {Suffix packets};
 \node[wh] at (0,-0.5) {}; \node[rule] at (0.21,-0.5) {Original packet (to be replaced)};
 \node[hook] at (5.4,-0.5) {}; \node[rule] at (5.55,-0.5) {Interception point (hook)};
 \draw[hand] (9.2,-0.5) -- (10.1,-0.5); \node[rule] at (10.2,-0.5) {Hand-off between phases};
\end{scope}
\draw[hand,preaction={draw=white,line width=4pt,dash pattern=}] ({\xP+4.65},9.25) -- ({\xP+4.65},7.09);
\node[lab,anchor=west] at ({\xP+4.7},8.4) {Prefix Tree};
\draw[hand] ({\xP+4.65},5.19) -- ({\xP+4.65},3.08);
\node[lab,anchor=west] at ({\xP+4.7},4.3) {Prefix Tree};
\draw[hand] (pool3.south) |- ({\xL-0.3},-3.3) -- ({\xL-0.3},11.63) -| (pool1.north);
\node[lab,rotate=90,fill=white] at ({\xL-0.3},4.17) {Next epoch: minimized queue};
\draw[hand] ({\xR-0.2},-1.75) -- ({\xR+0.3},-1.75) -- ({\xR+0.3},11.15) -- ({\xR-0.2},11.15);
\node[lab,rotate=-90,fill=white] at ({\xR+0.3},4.17) {Next epoch: trimmed Prefix Tree};
\end{tikzpicture}%
		}
		\caption{Overview of InterceptFuzz. Rows show the phases of one epoch, columns the components involved. Dashed violet arrows mark hand-offs between phases. Whenever the target runs, (1)~the fuzz entry passes the seed to the InterceptFuzz core, which (2)--(3)~looks up the prefix of the seed's State~ID in the Prefix Tree, and (4)~at each hook call replaces the intercepted packet with the next prefix packet or the payload, passing the suffix through unchanged, as detailed in Fig.~\ref{fig:running-example}. (5)~In Phase~I only, the payload and suffix extend the Prefix Tree, and each new edge is exported as a seed. In Phase~II, the engine mutates the entire seed, including the State~ID (the example shows an unchanged ID). Phase~III minimizes the engine queue by replaying every seed (IIIa) and then trims the Prefix Tree and renumbers the State~IDs once (IIIb).}
		\label{fig:architecture}
	\end{figure*}
	
	InterceptFuzz executes in iterative cycles called epochs, as illustrated in Figure~\ref{fig:architecture}. An external driver script orchestrates each epoch by running three operational phases strictly in sequence, with Phase III's output re-seeding Phase I of the next epoch to close the loop.
	
	All three phases operate over a shared state representation: a Prefix Tree that records the target's observed protocol behavior, together with a compact seed encoding that lets the underlying stateless engine address specific states within that tree. Section~\ref{sec:prefix-tree} defines both formally. The remaining phases build on this shared model as follows.
	
	\textbf{\textit{Phase I: Prefix Tree Extension}} builds and expands the Prefix Tree by replaying the seeds retained in the previous epoch and recording the traffic they produce, and exports the newly discovered transitions as seeds for the underlying stateless engine.
	
	\textbf{\textit{Phase II: Stateful Fuzzing and Dynamic Routing}} is where the actual fuzzing takes place. The stateless engine mutates the seeds exported by Phase I, and for each resulting input, InterceptFuzz first routes the target to the correct state before delivering it — effectively performing stateful fuzzing.
	
	\textbf{\textit{Phase III: Minimization and Trimming}} consolidates the epoch's results: the stateless engine's output queue is minimized, and the Prefix Tree is trimmed, aiming to mitigate the tree's unbounded growth without sacrificing the quality of the seeds it retains.
	
	\textbf{Epoch schedules.} The number of epochs and the length of each epoch in a fuzzing campaign are hyperparameters and can be chosen depending on the target's complexity. InterceptFuzz supports two schedules for distributing the fuzzing time of a campaign, $T_{\mathrm{total}}$, across $N$ epochs. Under the \emph{uniform} schedule, every epoch receives the same Phase~II budget of $T_{\mathrm{total}}/N$. Under the \emph{linear} schedule, the budget grows with the epoch index: epoch~$i \in \{1, \dots, N\}$ receives $i \cdot T$, where $T = T_{\mathrm{total}} / \sum_{j=1}^{N} j = 2\,T_{\mathrm{total}} / (N(N+1))$. The linear schedule builds on the observation that the Prefix Tree changes most during the first epochs, when many new transitions are still being discovered. Short early epochs pass the newly found seeds to Phase~I sooner, so that the tree is extended quickly, whereas the long later epochs, in which the tree grows only slowly, give the stateless engine more time to explore the accumulated states in depth. The uniform schedule, in contrast, refreshes the Prefix Tree at regular intervals throughout the campaign. Section~\ref{sec:eval:schedules} compares both schedules.

	The remainder of this section is organized as follows.
	Section~\ref{sec:prefix-tree} formally defines the Prefix Tree state model and the seed encoding used to address it. Section~\ref{sec:interception} describes InterceptFuzz's interception mechanism, the substrate shared by all three phases. Sections~\ref{sec:phase1}--\ref{sec:phase3} then detail each phase in turn.
	
	\subsection{State Model: The Prefix Tree}
	\label{sec:prefix-tree}
	
	Prior to describing the individual operational phases, this section formally establishes the state representation shared across all three: the Prefix Tree, its corresponding seed encoding, as well as the seed replay mechanism and resulting suffixes used to navigate and expand the model. Throughout this section, a \emph{packet} denotes a single protocol message passing through an interception point.
	
	\textbf{Prefix Tree.} InterceptFuzz models the target's observed execution histories as a rooted, directed tree. Each node denotes an abstract execution point of the target, reachable from the root $S_1$ by a unique sequence of intercepted packets. Nodes themselves carry no semantic meaning, while every transition is labeled with the concrete payload of the intercepted packet. For example, in the path $S_1 \xrightarrow{A} S_2 \xrightarrow{B} S_3$, replaying packet $A$ from $S_1$ deterministically reaches $S_2$, and replaying $B$ from $S_2$ reaches $S_3$. For state $S_i$, the sequence of packets labeling the unique path from $S_1$ to $S_i$ is called its \emph{prefix}, and $S_i$ together with all of its descendants its \emph{subtree}.
	
	\textbf{Node Identity.} InterceptFuzz ensures that each node's outgoing edges are labeled by pairwise-distinct packets: encountering an observed packet $P$ on an existing edge $S_i \xrightarrow{P} S_j$ reuses child $S_j$ rather than creating a duplicate branch. Consequently, any packet sequence identifies at most one path from the root, which guarantees deterministic routing during seed replay, a mechanism defined later in this section.
	InterceptFuzz intentionally adopts a coarse, history-based node identity instead of semantic state tracking, as recovering true equivalence across execution points would require target-specific internal knowledge that InterceptFuzz does not assume. The trade-off is potential unbounded tree growth, as syntactically distinct yet semantically redundant paths accumulate across epochs. Section~\ref{sec:phase3} describes how Phase~III bounds this growth through minimization and trimming.
	
	\textbf{Seed Encoding.} To let the underlying stateless engine address a specific node in the Prefix Tree, InterceptFuzz serializes every seed as a \texttt{[State~ID][payload]} pair: a 4-byte, big-endian State~ID identifying a node $S$, followed by the packet payload to be delivered once the target reaches $S$. State~IDs start at $1$ for the root $S_1$, and the ID $0$ is reserved for the null seed described in Section~\ref{sec:phase1}. The \texttt{[State~ID][payload]} pair is the sole interface between the Prefix Tree model and the underlying stateless engine: the engine only ever sees and mutates flat byte buffers in that format, while InterceptFuzz is responsible for resolving a State~ID back to a concrete node and for keeping the ID space consistent across epochs, as described in Section~\ref{sec:phase3}.
	
\begin{figure}[t]
		\centering
		\resizebox{\columnwidth}{!}{%
		\begin{tikzpicture}[
		  font=\footnotesize,
		  st/.style={circle,draw,thick,minimum size=6.5mm,inner sep=0pt},
		  newst/.style={st,draw=orange!80!black,fill=orange!8},
		  e/.style={-{Stealth[length=2mm]},thick},
		  lbl/.style={midway,above,inner sep=1.5pt},
		  pk/.style={rectangle,draw,thick,minimum width=7mm,minimum height=5mm,inner sep=1pt},
		  pre/.style={pk,fill=gray!15},
		  pay/.style={pk,fill=orange!25,draw=orange!80!black},
		  suf/.style={pk,fill=blue!10,draw=blue!60!black},
		  brc/.style={decorate,decoration={brace,amplitude=4pt,mirror},thick},
		]
		\node[st] (a0) at (0,0) {$S_1$};
		\node[st] (a1) at (1.6,0) {$S_2$};
		\node[st,fill=gray!15] (a2) at (3.2,0) {$S_3$};
		\node[st] (a3) at (4.8,0) {$S_4$};
		\draw[e] (a0) -- node[lbl]{$A$} (a1);
		\draw[e] (a1) -- node[lbl]{$B$} (a2);
		\draw[e] (a2) -- node[lbl]{$C$} (a3);
		\node[draw,thick,rounded corners=1pt,fill=white,font=\ttfamily\footnotesize] (seed) at (6.6,0.9) {[3][D]};
		\node[above=0pt of seed,font=\scriptsize] {seed};
		\draw[-{Stealth[length=2mm]},dashed,thick] (seed.west) to[out=180,in=60] node[above,font=\scriptsize,pos=0.45]{State~ID} (a2.north east);
		
		\node[font=\scriptsize,anchor=east] at (-0.2,-1.3) {hook call};
		\foreach \i/\x in {1/0,2/1.1,3/2.2,4/3.3,5/4.4} \node[font=\scriptsize] at (\x+0.4,-1.3) {\#\i};
		\node[pre] (h1) at (0.4,-1.85) {$A$};
		\node[pre] (h2) at (1.5,-1.85) {$B$};
		\node[pay] (h3) at (2.6,-1.85) {$D$};
		\node[suf] (h4) at (3.7,-1.85) {$E$};
		\node[suf] (h5) at (4.8,-1.85) {$F$};
		\foreach \a/\b in {h1/h2,h2/h3,h3/h4,h4/h5} \draw[e,gray] (\a) -- (\b);
		\node[font=\scriptsize,anchor=east] at (-0.2,-1.85) {packet};
		\draw[brc] ($(h1.south west)+(0,-0.08)$) -- ($(h2.south east)+(0,-0.08)$) node[midway,below=4pt,align=center,font=\scriptsize]{prefix\\injected};
		\draw[brc] ($(h3.south west)+(0,-0.08)$) -- ($(h3.south east)+(0,-0.08)$) node[midway,below=4pt,align=center,font=\scriptsize]{payload\\injected};
		\draw[brc] ($(h4.south west)+(0,-0.08)$) -- ($(h5.south east)+(0,-0.08)$) node[midway,below=4pt,align=center,font=\scriptsize]{suffix\\observed};
		\node[anchor=west,font=\small\bfseries] at (-1.6,1.2) {(a)};
		\node[anchor=west] at (-0.95,1.2) {Seed replay of \texttt{[3][D]}};
		
		\begin{scope}[yshift=-4.6cm]
		\node[anchor=west,font=\small\bfseries] at (-1.6,1.0) {(b)};
		\node[anchor=west] at (-0.95,1.0) {Prefix Tree after Phase~I};
		\node[st] (b0) at (0,0) {$S_1$};
		\node[st] (b1) at (1.6,0) {$S_2$};
		\node[st] (b2) at (3.2,0) {$S_3$};
		\node[st] (b3) at (4.8,0) {$S_4$};
		\draw[e] (b0) -- node[lbl]{$A$} (b1);
		\draw[e] (b1) -- node[lbl]{$B$} (b2);
		\draw[e] (b2) -- node[lbl]{$C$} (b3);
		\node[newst] (b4) at (3.2,-1.4) {$S_5$};
		\node[newst] (b5) at (4.8,-1.4) {$S_6$};
		\node[newst] (b6) at (6.4,-1.4) {$S_7$};
		\draw[e,orange!80!black] (b2) -- node[midway,left,inner sep=2pt]{$D$} (b4);
		\draw[e,blue!60!black] (b4) -- node[lbl]{$E$} (b5);
		\draw[e,blue!60!black] (b5) -- node[lbl]{$F$} (b6);
		\node[font=\ttfamily\scriptsize,text=orange!60!black,anchor=east] at (2.75,-0.7) {[3][D]};
		\node[font=\ttfamily\scriptsize,text=blue!50!black,below=1pt] at ($(b4)!0.5!(b5)+(0,-0.12)$) {[5][E]};
		\node[font=\ttfamily\scriptsize,text=blue!50!black,below=1pt] at ($(b5)!0.5!(b6)+(0,-0.12)$) {[6][F]};
		\node[font=\scriptsize,align=left,anchor=west] at (5.3,-0.55) {new edges and\\exported seeds};
		\end{scope}
		\end{tikzpicture}%
		}
		\caption{Running example. (a) Seed replay of \texttt{[3][D]}: InterceptFuzz injects the recorded prefix $\langle A,B\rangle$ of $S_3$ at the first two hook calls, delivers the payload $D$ at the third, and passively observes the suffix $\langle E,F\rangle$. (b) Phase~I adds the payload and the suffix to the Prefix Tree and exports one seed per new edge.}
		\label{fig:running-example}
	\end{figure}
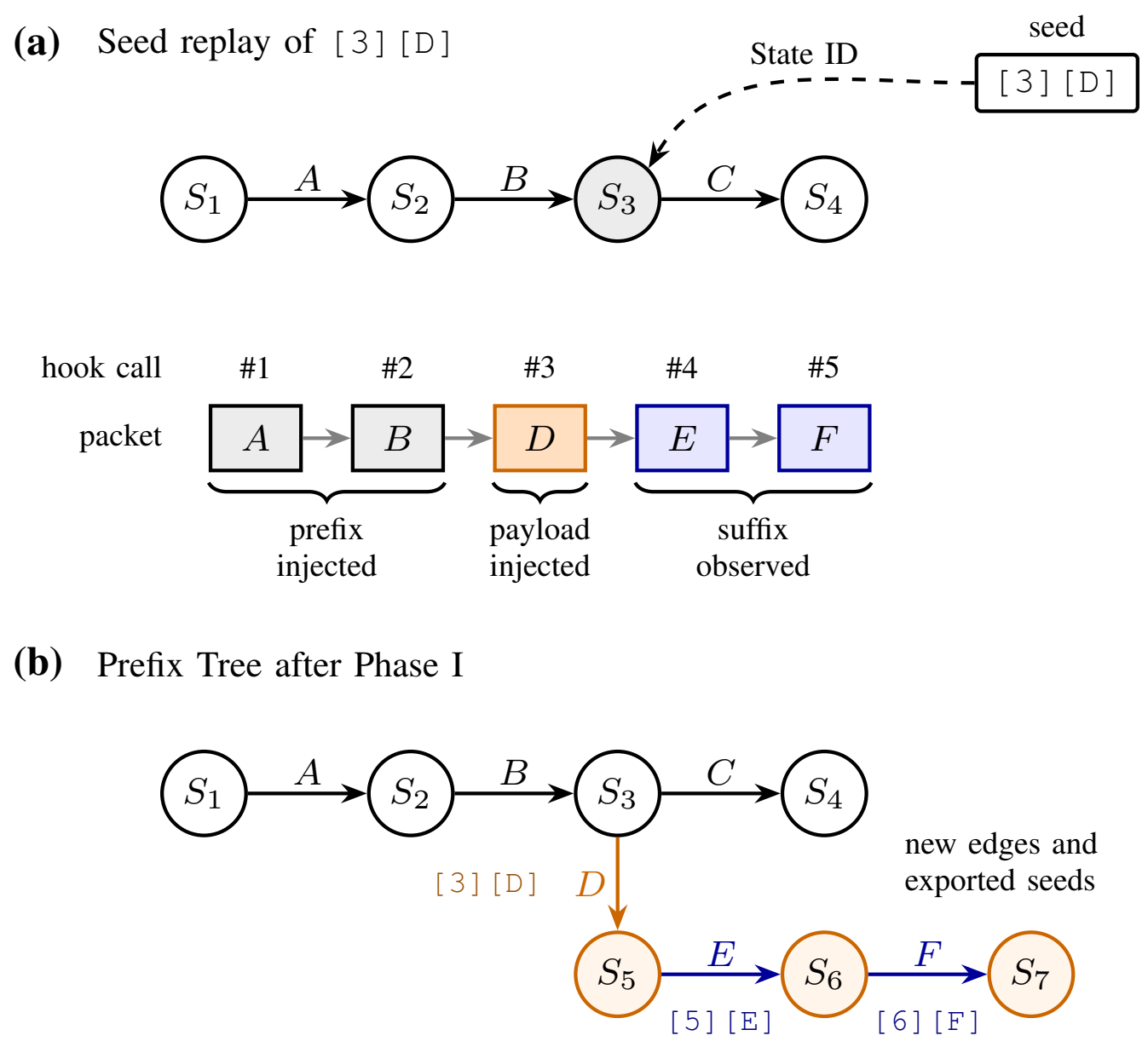

	\textbf{Seed Replay.} Executing a seed---routing the target to a designated prefix state before delivering a payload---is the core operation shared by Phase~I and Phase~II, termed \emph{seed replay}. InterceptFuzz implements this mechanism via its interception hook. Specifically, on each successive hook trigger, the InterceptFuzz middleware injects the next packet along the path to the target node. Once the prefix is exhausted, it delivers the seed's payload. For example, consider the seed \texttt{[3][D]} evaluated against the Prefix Tree $S_1 \xrightarrow{A} S_2 \xrightarrow{B} S_3 \xrightarrow{C} S_4$. The state identifier \texttt{3} corresponds to node $S_3$, which has the prefix sequence $\langle A, B \rangle$. InterceptFuzz sequentially injects $A$ on the first hook trigger and $B$ on the second to reach $S_3$, and finally injects payload $D$ on the third trigger upon prefix completion, as illustrated in Figure~\ref{fig:running-example}a. Under the determinism assumption, replaying a prefix triggers the hook at exactly the same packet boundaries as during recording, so InterceptFuzz implements no divergence handling.
	
	\textbf{Suffix.} The sequence of packets emitted by the target after seed replay completes constitutes that seed's \emph{suffix}. Whereas the prefix replays recorded traffic and the payload originates from the seed, the suffix is neither injected nor altered by InterceptFuzz but passively observed, reflecting the target's autonomous protocol behavior at the reached execution state. Continuing the previous example, the suffix of seed \texttt{[3][D]} comprises any packets the target emits following the delivery of $D$, such as $\langle E, F \rangle$ in Figure~\ref{fig:running-example}a.
	
	\textbf{Relation to AFLNet.} Seed replay is structurally close to the input model of AFLNet, which also replays the messages that lead to a selected state unchanged and mutates the messages after them. The two approaches differ in three respects. First, they differ in what they consider a state. AFLNet identifies a state by the response code of the target, which requires a protocol whose responses carry such codes, or a protocol-specific dissector that derives them. In InterceptFuzz, the state model is much more generic and follows from the coverage feedback, i.e., new states are added to the Prefix Tree based on the inputs retained in the corpus of the underlying engine. InterceptFuzz therefore requires no knowledge of the protocol to identify states and is not tied to any particular protocol. Second, they differ in how a state is selected. AFLNet adds a dedicated state scheduler to the engine, which first selects a state and then a seed that reaches it. InterceptFuzz instead encodes the state in the input as a State~ID, so that selecting a seed also selects its state. State selection is thus left to the seed scheduling of the unmodified engine, and InterceptFuzz benefits from any improvement to it. Third, AFLNet acts as a client that sends its messages to a server over a socket, whereas InterceptFuzz delivers packets at an in-memory interception point, as described next.
	
	\subsection{Interception and Execution Model}
	\label{sec:interception}
	To interact with in-process targets that communicate without OS-level sockets, InterceptFuzz utilizes a minimal source-level integration strategy. InterceptFuzz exposes a single interception hook function, which can be inserted wherever traffic is to be fuzzed---typically at the points where the target produces or consumes packets. In the simplest case, each insertion is a single line that calls the hook, \texttt{size\_t InterceptFuzz::intercept(buf, buf\_len, buf\_max\_len)}, on the packet buffer. For targets with strict memory ownership models (e.g., read-only packet structures), a variant of the hook, \texttt{InterceptFuzz::intercept\_copy}, returns the resulting packet as a copy, and a few additional lines write it back into the execution flow. When the hook is invoked from several code locations or on behalf of several simulated nodes, InterceptFuzz treats all invocations as a single, globally ordered sequence in which every intercepted packet becomes an edge of the Prefix Tree. Determinism guarantees that this order is identical across executions. To fuzz only part of the traffic, e.g., packets sent by a single node acting as a malicious peer, the developer can simply call the hook inside an \texttt{if} statement on that condition.
	
	The \texttt{InterceptFuzz::intercept} hook introduced above functions as InterceptFuzz's exclusive gateway for both data observation and manipulation: it inspects or overwrites the packet in \texttt{buf} and returns the resulting payload length. When invoked, the InterceptFuzz library evaluates the target buffer against the current execution phase. During the first epoch, when the Prefix Tree is still empty, the hook passively observes and records benign packets, as described in Section~\ref{sec:phase1}. During Phase~I expansion and Phase~II fuzzing alike, the hook executes seed replay and leaves the resulting suffix untouched. The two phases differ only in what is recorded: Phase~I adds the delivered payload and the resulting suffix to the Prefix Tree as newly discovered transitions, while Phase~II records neither. Finally, to preserve the memory safety of the simulation, InterceptFuzz truncates all injected payloads to \texttt{buf\_max\_len}, the maximum payload size that the developer allows at that particular hook call. This value therefore depends on the target and on where the hook is placed.
	
	The hook thus governs what happens within a single execution. Starting executions and resetting the target between them is left entirely to the underlying fuzzing engine, and InterceptFuzz implements no reset mechanism of its own. Whether the engine isolates executions with a fork server, as AFL++ does, or relies on an in-process reset harness, as libFuzzer does, InterceptFuzz works the same way. Likewise, coverage feedback is provided entirely by the engine: the target, including all simulated nodes, is compiled with the engine's instrumentation, so the engine decides which inputs to retain exactly as it would for a stateless target.
	
	\subsection{Phase I: Prefix Tree Extension}
	\label{sec:phase1}
	
	The task of Phase I is to populate the Prefix Tree with new transitions. It takes as input the minimized queue of seeds handed off by the previous epoch's Phase III and performs seed replay on each seed in turn. Because the prefix is replayed from the Prefix Tree, it is already recorded there, so InterceptFuzz only adds the delivered payload and the resulting suffix to the tree. At the end of this process, the newly discovered transitions are exported as initial seeds, using the \texttt{[State~ID][payload]} encoding, for the underlying stateless engine to be mutated in Phase II. Concretely, each newly added edge $S_i \xrightarrow{P} S_j$ yields the seed \texttt{[$i$][$P$]}, so that the packet observed at $S_i$---benign or previously mutated---becomes the starting point for mutations delivered at that state. Since the replayed queue has already been minimized in Phase~III, the number of new edges, and hence of exported seeds, is bounded by the number of retained seeds and the lengths of their suffixes. For example, replaying \texttt{[3][D]} on the Prefix Tree $S_1 \xrightarrow{A} S_2 \xrightarrow{B} S_3 \xrightarrow{C} S_4$ adds the edge $S_3 \xrightarrow{D} S_5$ and, for a suffix $\langle E, F \rangle$, the edges $S_5 \xrightarrow{E} S_6$ and $S_6 \xrightarrow{F} S_7$, yielding the seeds \texttt{[3][D]}, \texttt{[5][E]}, and \texttt{[6][F]}, as shown in Figure~\ref{fig:running-example}b.
	
	In the first epoch, when the Prefix Tree is empty and no seeds yet exist, InterceptFuzz uses a reserved null seed --- State~ID~0 paired with an empty payload --- which triggers an iteration where InterceptFuzz simply observes the target's benign traffic from the very beginning, building the initial Prefix Tree skeleton from it.
	
	\subsection{Phase II: Stateful Fuzzing and Dynamic Routing}
	\label{sec:phase2}
	
	Once Phase~I has extended the Prefix Tree and exported the new seeds, InterceptFuzz transitions to Phase~II, where the underlying stateless engine mutates these seeds while InterceptFuzz routes the target to the state each input addresses. Because the engine operates on inputs encoded as \texttt{[State~ID][payload]}, InterceptFuzz must interpret each mutated input before dispatching execution.
	
	This decoding takes place in InterceptFuzz's fuzz-entry function---the harness callback (e.g., an \texttt{LLVMFuzzerTestOneInput}-style entry point) invoked directly by the stateless engine. The function extracts the initial 4-byte State~ID before the target harness is invoked. If the input is shorter than 4 bytes or specifies an ID that does not map to an existing node in the Prefix Tree, the callback returns immediately, making invalid inputs cheap to reject.
	For valid IDs, InterceptFuzz performs seed replay to route the target to the designated state and then delivers the rest of the mutated input as the payload, truncated to \texttt{buf\_max\_len}. Each execution thus mutates only a single packet, as the prefix is replayed exactly as recorded. However, inputs that increase coverage are retained in the engine's queue and replayed in the next epoch's Phase~I, where their mutated payloads become edges of the Prefix Tree. As a result, a trace can accumulate multiple mutated packets over successive epochs.
	
	The length of Phase~II in each epoch is set by the epoch schedule introduced in Section~\ref{sec:overview}: by default, it grows linearly, so that early epochs return new seeds to Phase~I quickly and later epochs explore the accumulated states longer. Furthermore, the State~ID remains deliberately exposed to mutation by the underlying fuzzing engine. Because the identifier occupies only a small fraction of the input buffer, it is mutated infrequently. Moreover, since IDs are assigned contiguously, most mutations of its high-order bytes yield an invalid ID that is rejected at negligible cost, as described above. When a mutation does produce another valid ID, the same payload is delivered at a different state, i.e., a packet observed in one protocol state is replayed in another. Such out-of-context packets exercise the target's handling of unexpected messages and thus benefit exploration.
	
	\subsection{Phase III: Minimization and Trimming}
	\label{sec:phase3}
	When a Phase~II fuzzing campaign concludes, the stateless engine yields a queue of interesting seeds---mutated \texttt{[State~ID][payload]} inputs that uncovered new branch coverage. However, directly incorporating every coverage-increasing input into the Prefix Tree would cause substantial redundancy and rapid growth of the tree. InterceptFuzz mitigates this growth through two complementary techniques.
	
	First, the fuzzer's output queue is distilled to the minimal corpus exercising the newly discovered coverage using the engine's corpus minimizer (e.g., \texttt{afl-cmin} for AFL++). Next, InterceptFuzz trims the Prefix Tree against this reduced queue: a node $S_i$ is pruned if and only if neither $S_i$ nor any of its descendants are referenced by any retained seed. Evaluating survival across the entire subtree ensures an ancestor is never removed while a descendant remains active, preserving the prefix path required for deterministic seed replay. Consequently, the root $S_1$ is never pruned as long as at least one valid node survives.
	
	Because trimming leaves holes in the State~ID sequence, the pass concludes by renumbering the nodes to restore contiguity: the root remains $S_1$, while the $N$ surviving non-root nodes are reassigned continuous IDs from $2$ to $N+1$. To keep the minimized queue consistent, InterceptFuzz rewrites each surviving seed's 4-byte State~ID field in place to match its node's updated ID. This relabeled queue then loops back into Phase~I to drive expansion in the next epoch.
	
	\section{Evaluation}
	\label{sec:evaluation}
	This section evaluates InterceptFuzz on the two single-process protocol simulators introduced in Section~\ref{sec:bg:simulation} against two baseline fuzzers, Concat and AFLNet, and answers the following research questions:
	\begin{itemize}
		\item \textbf{RQ1 (Coverage):} Does InterceptFuzz cover more code of stateful single-process targets than the baselines?
		\item \textbf{RQ2 (Bug finding):} Does InterceptFuzz expose more crash signatures than the baselines, and does it expose them more reliably across campaigns?
		\item \textbf{RQ3 (State model):} How does the Prefix Tree evolve across epochs, and how effectively does Phase~III keep its growth in check?
		\item \textbf{RQ4 (Epoch settings):} How sensitive is InterceptFuzz to the number of epochs and to the epoch schedule?
		\item \textbf{RQ5 (Integration effort):} How much manual effort does it take to integrate InterceptFuzz into a new target?
	\end{itemize}
	
	\subsection{Experimental Setup}
	\label{sec:eval:setup}
	\textbf{Targets.} The evaluation uses three targets built on the two single-process simulators: the Nexus test framework of OpenThread and the Wi-Fi and LTE modules of ns-3. The evaluation uses OpenThread at commit \texttt{01bce95b} and ns-3 at commit \texttt{be3db7b6}. The targets cover two kinds of software. As discussed in Section~\ref{sec:bg:simulation}, ns-3 is a simulator used mainly in networking research, whereas OpenThread is an open-source protocol stack, which is also used in commercial Thread products. The targets also differ in how much code their scenarios can reach beyond the benign trace, from a few hundred branches on ns-3 LTE to several thousand on OpenThread Nexus. In each target, the harness runs a minimal scenario in which one node attaches to, or associates with, another, so that the fuzzed traffic covers the procedure from its first message onward:
	\begin{itemize}
		\item \emph{OpenThread Nexus}: a leader forms the network, and a Thread device then joins it as a full end device through the MLE attach procedure. The benign run contains four messages: Parent Request, Parent Response, Child ID Request, and Child ID Response.
		\item \emph{ns-3 Wi-Fi}: one station associates with one access point within 300\,ms of simulated time. The benign run contains seven frames: a Probe Request and Probe Response, an Association Request and Association Response, and three Beacons.
		\item \emph{ns-3 LTE}: one user equipment (UE) attaches to one eNB, and the run ends as soon as the RRC connection is established or times out after 500\,ms of simulated time. As ns-3 does not implement the NAS logic, the fuzzed exchange consists of the RRC messages between the UE and the eNB. In the benign run, these are RRC Connection Request, RRC Connection Setup, and RRC Connection Setup Complete.
	\end{itemize}
	In all three targets, every intercepted packet can be mutated, regardless of which node sent it. All executions are deterministic: every input is executed in a freshly forked process, and at the start of each execution every harness reseeds the C standard library's pseudo-random number generator (\texttt{srand}).
	
	\textbf{Hook placement.} In each target, the hook is placed at a point where outgoing packets are available in serialized form, before they reach the receiving node:
	\begin{itemize}
		\item \emph{OpenThread Nexus}: a single hook is placed in the simulated radio of the Nexus platform, where it intercepts every IEEE~802.15.4 frame on its way to the receiving node and may mutate the entire frame. The frame check sequence is set to a fixed value, as the fuzzing build does not verify it. Besides MLE messages, this includes messages of the Thread Management Framework (TMF), which nodes exchange over CoAP once they are attached and which do not occur in the benign run.
		\item \emph{ns-3 Wi-Fi}: a single hook is placed in the MAC transmit queue (\texttt{WifiMacQueue::Enqueue}), through which every frame of both the access point and the station passes. The hook mutates only the frame body and keeps the MAC header, so that frames still reach their receiver. Frames without a body, such as acknowledgments, are not intercepted.
		\item \emph{ns-3 LTE}: every RRC message is serialized in a dedicated send function of the UE or the eNB, such as \texttt{DoSendRrcConnectionRequest}. Each of the 11 send functions receives a one-line call to a small helper, which copies the serialized message into a mutable buffer and invokes the hook.
	\end{itemize}
	Table~\ref{tab:targets} summarizes the three targets.
	
	\begin{table}[t]
		\centering
		\caption{Evaluation targets and integration effort.}
		\label{tab:targets}
		\setlength{\tabcolsep}{2pt}
		\begin{tabular}{lccccc}
			\hline
			\textbf{Target} & \textbf{Version} & \textbf{Benign} & \textbf{Hook} & \textbf{Target} & \textbf{Harness} \\
			 & & \textbf{packets} & \textbf{calls} & \textbf{lines} & \textbf{lines} \\
			\hline
			ns-3 LTE & \texttt{be3db7b6} & 3 & 11 & 21 & 59 \\
			ns-3 Wi-Fi & \texttt{be3db7b6} & 7 & 1 & 11 & 48 \\
			OpenThread Nexus & \texttt{01bce95b} & 4 & 1 & 6 & +19/$-$47$^*$ \\
			\hline
		\end{tabular}
		\\[2pt]
		\parbox{\columnwidth}{\footnotesize $^*$Added/removed lines with respect to an existing OpenThread integration test.}
	\end{table}
	
	\begin{table*}[t]
		\centering
		\caption{Final code coverage, median and [min--max] over the campaigns, and gain of the median over the benign trace ($\Delta$). Execs/s: executions per second, median [min--max]. InterceptFuzz uses five epochs under the linear schedule. The highest median coverage per target is in bold. The last two columns compare the branch coverage of InterceptFuzz with that of each baseline (two-sided Mann-Whitney $U$ test and Vargha-Delaney $\hat{A}_{12}$, the probability that a run of InterceptFuzz covers more branches than a run of the baseline).}
		\label{tab:coverage}
		\setlength{\tabcolsep}{2.5pt}
		\begin{tabular}{lrrrrrrrcc}
			\hline
			& & \multicolumn{2}{c}{\textbf{Branches}} & \multicolumn{2}{c}{\textbf{Lines}} & \multicolumn{2}{c}{\textbf{Functions}} & \multicolumn{2}{c}{\textbf{IF vs.\ baseline}} \\
			\textbf{Fuzzer} & Execs/s & Covered & $\Delta$ & Covered & $\Delta$ & Covered & $\Delta$ & $p$ & $\hat{A}_{12}$ \\
			\hline
			\multicolumn{10}{l}{\textit{OpenThread Nexus (10 runs)}} \\
			\quad Benign trace & --- & 2,634 & --- & 11,162 & --- & 3,312 & --- & &  \\
			\quad AFLNet & 308 {\scriptsize [295--308]} & 4,323 {\scriptsize [4,219--4,555]} & +1,689 & 14,350 {\scriptsize [14,056--14,973]} & +3,188 & 3,850 {\scriptsize [3,789--3,991]} & +538 & $<$0.001 & 1.00 \\
			\quad Concat & 952 {\scriptsize [685--1,067]} & 6,603 {\scriptsize [6,248--6,790]} & +3,969 & 18,666 {\scriptsize [17,972--19,203]} & +7,504 & 4,613 {\scriptsize [4,474--4,732]} & +1,301 & 0.001 & 0.94 \\
			\quad InterceptFuzz & 723 {\scriptsize [699--752]} & \textbf{6,915} {\scriptsize [6,715--7,116]} & \textbf{+4,281} & \textbf{19,265} {\scriptsize [18,953--19,690]} & \textbf{+8,103} & \textbf{4,727} {\scriptsize [4,676--4,834]} & \textbf{+1,415} &  &  \\
			\multicolumn{10}{l}{\textit{ns-3 Wi-Fi (10 runs)}} \\
			\quad Benign trace & --- & 40,496 & --- & 46,106 & --- & 13,567 & --- & &  \\
			\quad AFLNet & 29 {\scriptsize [27--38]} & 40,729 {\scriptsize [40,699--40,863]} & +233 & 46,315 {\scriptsize [46,238--46,503]} & +209 & 13,599 {\scriptsize [13,587--13,628]} & +32 & $<$0.001 & 1.00 \\
			\quad Concat & 29 {\scriptsize [28--38]} & 41,251 {\scriptsize [41,143--41,389]} & +755 & 46,794 {\scriptsize [46,669--47,084]} & +688 & 13,646 {\scriptsize [13,639--13,689]} & +79 & 0.011 & 0.84 \\
			\quad InterceptFuzz & 34 {\scriptsize [33--44]} & \textbf{41,363} {\scriptsize [41,054--41,459]} & \textbf{+867} & \textbf{47,041} {\scriptsize [46,585--47,106]} & \textbf{+935} & \textbf{13,687} {\scriptsize [13,631--13,696]} & \textbf{+120} &  &  \\
			\multicolumn{10}{l}{\textit{ns-3 LTE (10 runs)}} \\
			\quad Benign trace & --- & 34,375 & --- & 40,955 & --- & 12,219 & --- & &  \\
			\quad AFLNet & 36 {\scriptsize [36--47]} & \textbf{34,989} {\scriptsize [34,820--35,035]} & \textbf{+614} & 41,997 {\scriptsize [41,768--42,038]} & +1,042 & \textbf{12,370} {\scriptsize [12,341--12,378]} & \textbf{+151} & 0.791 & 0.46 \\
			\quad Concat & 29 {\scriptsize [28--38]} & 34,900 {\scriptsize [34,886--34,927]} & +525 & 41,947 {\scriptsize [41,904--42,038]} & +992 & 12,349 {\scriptsize [12,345--12,352]} & +130 & 0.225 & 0.67 \\
			\quad InterceptFuzz & 69 {\scriptsize [64--94]} & 34,922 {\scriptsize [34,913--35,113]} & +547 & \textbf{42,021} {\scriptsize [41,994--42,216]} & \textbf{+1,066} & 12,352 {\scriptsize [12,352--12,385]} & +133 &  &  \\
			\hline
		\end{tabular}
	\end{table*}

	\textbf{Baselines.} InterceptFuzz is compared against two baselines, both of which use the same interception points as InterceptFuzz.
	\begin{itemize}
		\item Concat: a concatenation baseline, implemented specifically for this evaluation, which realizes the seed concatenation strategy of Section~\ref{sec:stateless_bypasses} on top of AFL++. A harness first splits each input into messages at a delimiter, and the $i$-th hook call then replaces the intercepted packet with the $i$-th message. Once all messages have been delivered, the remaining packets pass through unchanged, and surplus messages are discarded.
		\item \emph{AFLNet}: the LibAFL reimplementation released with Fizzle~\cite{fizzle} (artifact commit \texttt{e1104bc}), whose socket communication is replaced by the interception points. AFLNet infers its state model from response codes, which the evaluated protocols do not provide. Therefore, for each target, a separate dissector was written. It maps every captured response to a code that identifies its message type, following the recommendation of the AFLNet authors to expose an explicit state identifier when responses carry no state information. In preliminary experiments, several dissectors that distinguish the responses at different levels of detail were tried, and the evaluation uses the one that performed best. On OpenThread Nexus, the size of mutated messages is additionally capped at the maximum IEEE~802.15.4 frame size of 127~bytes, so that mutations are not spent on bytes that a frame cannot carry.
	\end{itemize}
	All three fuzzers therefore share one delivery mechanism, so that differences in the results stem from the fuzzers themselves rather than from how their inputs reach the target.
	
	\textbf{Initial seeds.} All three fuzzers start from the same benign traffic of an unfuzzed run of each target. Because the targets are deterministic, a single run suffices. Both baselines natively expect this run to be recorded beforehand. For Concat, the recorded packets are stored as a single delimiter-separated input. For AFLNet, a bootstrapping step, which is not part of the Fizzle release, was added for this evaluation. It records the run at the start of each campaign and stores it as a message sequence. InterceptFuzz records the run without any such extension, by replaying the null seed into its Prefix Tree in the first epoch. The initial seeds of all three fuzzers thus contain the same packets.
	
	\textbf{Configuration.} InterceptFuzz and Concat both use AFL++~\cite{aflplusplus} 4.32c (commit \texttt{247e8241}) as the underlying engine. Each campaign runs for 24~hours on OpenThread Nexus and for 10~hours on the two ns-3 targets, whose fuzzed scenarios are smaller, and is repeated 10 times per fuzzer and target. By default, InterceptFuzz divides each campaign into five epochs under the linear schedule of Section~\ref{sec:overview}, so that the Phase~II budgets grow from 1.6 to 8~hours on OpenThread Nexus (1.6, 3.2, 4.8, 6.4, and 8~hours) and from 40~minutes to 3.3~hours on the ns-3 targets (40, 80, 120, 160, and 200~minutes). Phases~I and~III run in addition to these budgets but take only 0.4\% of the campaign time on OpenThread Nexus and 0.3--1.1\% on the ns-3 targets. Section~\ref{sec:eval:schedules} compares this default with three alternative settings on OpenThread Nexus: five epochs under the uniform schedule, and ten epochs under the linear and under the uniform schedule. All experiments run on an HP EliteBook 850 G8 laptop with an Intel Core i7-1165G7 CPU (4 cores, 8 hardware threads) and 16\,GB of RAM, running Ubuntu 22.04. Each campaign runs in its own Docker container with two dedicated hardware threads and a single fuzzer instance, and four campaigns run in parallel.
	
	\textbf{Metrics.} The comparison with the baselines uses two metrics: code coverage and the distinct crash signatures that each fuzzer exposes. Code coverage is measured by replaying the final queue of each campaign on a separate gcov build of the target. Branch coverage is the primary metric. Following common recommendations for fuzzer evaluation~\cite{klees2018evaluating,arcuri2011practical}, the results report absolute numbers of covered branches, and differences are tested with the Mann-Whitney $U$ test and the Vargha-Delaney $\hat{A}_{12}$ effect size. The fuzzing builds are compiled without AddressSanitizer (ASan), as its instrumentation slows down every execution and would reduce the throughput of all three fuzzers. Crashes during fuzzing are therefore limited to failures that terminate the uninstrumented binary, such as assertion failures and segmentation faults. To expose memory errors that do not crash it, the non-minimized queue of each campaign is afterwards replayed on an ASan build, so that no input is lost to minimization. Note, this setup can miss some memory errors from the inputs that are not kept in the queue, for example inputs that corrupt memory without reaching new coverage. Therefore, this approach is a compromise between throughput and the detection of memory errors. But, as it affects all three fuzzers in the same way, the comparison nevertheless stays fair. Finally, crashes are deduplicated into crash signatures, given by the assertion message or, for sanitizer reports, by the function in which the fault occurs, and every crash signature was confirmed manually to be reproducible.
	
	\subsection{Attacker Model}
	\label{sec:eval:attacker}
	Whether a crash found by fuzzing is a vulnerability depends on who can send the messages that trigger it. This section therefore defines the attacker for each target and discusses how the fuzzing setup relates to this attacker.

	\textbf{OpenThread.} The attacker is a malicious device that already possesses the shared network key and can therefore communicate with the other devices of the Thread network. It attacks a node during the MLE attach procedure by sending it malformed messages. Since the attacker holds the key, it can encrypt and authenticate any message it sends. The fuzzing builds define \texttt{FUZZING\_BUILD\_MODE\_UNSAFE\_FOR\_PRODUCTION}, the standard macro for fuzzing builds, under which OpenThread skips the decryption and integrity check of received MLE messages. This only removes a step that the attacker would pass anyway. How the key is obtained during commissioning is out of scope.

	\textbf{ns-3.} ns-3 is a research simulator and is not deployed in production, so the crashes found in ns-3 LTE and ns-3 Wi-Fi are not vulnerabilities in practice. If the protocol code of ns-3 were used on a real device, the corresponding attacker would be a device within radio range of the victim: for LTE, a rogue base station attacking a UE or a malicious UE attacking a base station, and for Wi-Fi, a rogue access point or a malicious station.

	\textbf{Mutations in both directions.} All three fuzzers mutate packets in both directions of the exchange, whereas a real attacker controls only the packets sent by its own device. A crashing input may therefore combine mutations of packets that the attacker cannot control, such as packets sent by the victim itself, and such a crash may not be reproducible against a real device. In general, deciding whether a given crash is reachable by the attacker is difficult, because a crash that requires a mutated packet from the victim on one path may still be reachable on another path that requires only packets from the attacker. Such crashes are nevertheless real security findings: the crashing node fails on a message that it can receive, and a later change to the protocol or to the code may make them reachable by an attacker. These crashes should therefore be fixed as well.

	\subsection{RQ1: Coverage}
	\label{sec:eval:coverage}
\begin{figure*}[t]
		\centering
		\begin{minipage}{\columnwidth}
			\centering
			\includegraphics[width=\columnwidth]{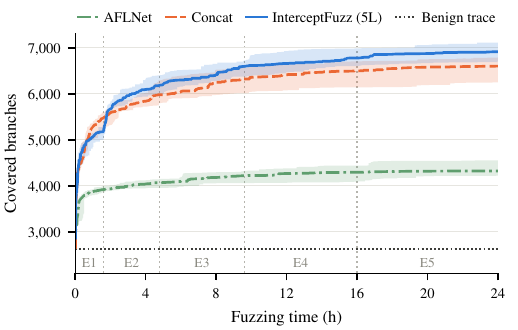}\\
			\footnotesize (a) OpenThread Nexus (24~hours)
		\end{minipage}\hfill
		\begin{minipage}{\columnwidth}
			\centering
			\includegraphics[width=\columnwidth]{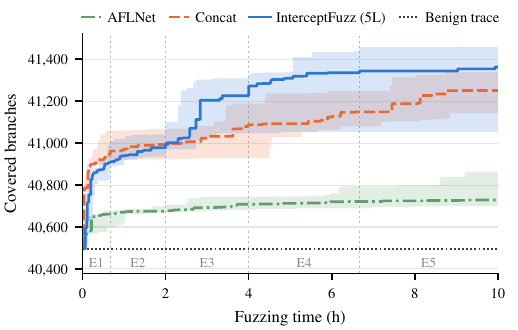}\\
			\footnotesize (b) ns-3 Wi-Fi (10~hours)
		\end{minipage}\\[1.5ex]
		\begin{minipage}{\columnwidth}
			\centering
			\includegraphics[width=\columnwidth]{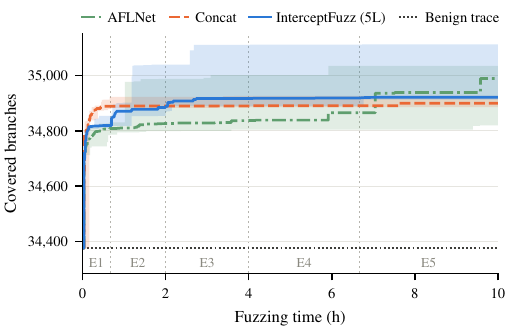}\\
			\footnotesize (c) ns-3 LTE (10~hours)
		\end{minipage}
		\caption{Branch coverage over time on (a) OpenThread Nexus, (b) ns-3 Wi-Fi, and (c) ns-3 LTE: median (line) and min--max (shaded band) over ten campaigns per fuzzer. The dark horizontal line marks the coverage of the benign trace without fuzzing, from which all fuzzers start at time zero. The light vertical lines mark the ends of the five epochs of InterceptFuzz (E1--E5) under the linear schedule.}
		\label{fig:covtime}
	\end{figure*}

	RQ1 compares the code coverage that the three fuzzers reach on each target. Table~\ref{tab:coverage} lists the final branch, line, and function coverage of each fuzzer, both in total and beyond the benign trace, which isolates the code reached only through fuzzing, together with the throughput of each fuzzer. Branch coverage is the primary metric, while line and function coverage show whether the differences in branch coverage carry over to coarser measures. The results are presented per target. Fig.~\ref{fig:covtime} shows how the branch coverage develops over the course of the campaigns.

	\subsubsection{OpenThread Nexus}
	InterceptFuzz performs the best, covering a median of 4,281 branches beyond the benign trace, against 3,969 for Concat and 1,689 for AFLNet. Both differences are statistically significant ($p = 0.001$ with $\hat{A}_{12} = 0.94$, and $p < 0.001$ with $\hat{A}_{12} = 1.00$), and every InterceptFuzz campaign covers more branches than the median Concat campaign. The same holds for lines and functions ($p \le 0.001$). Over time, Concat covers more branches than InterceptFuzz for most of the first epoch, in which InterceptFuzz fuzzes only the states of the benign trace, as Fig.~\ref{fig:covtime}a shows. InterceptFuzz overtakes it after about 1.8~hours, shortly after the first extension of its Prefix Tree, and then stays ahead for the rest of the campaign. AFLNet remains far below both fuzzers throughout.

	The coverage analysis showed that most of the additional branches of InterceptFuzz lie in the handling of TMF messages, which the nodes exchange only after the attach and which the benign run does not contain. InterceptFuzz first drives the nodes into exchanging such messages and then fuzzes them in the states recorded in its Prefix Tree. A more detailed breakdown of the coverage can be seen in Appendix~\ref{app:coverage}.

	\subsubsection{ns-3}
	On ns-3 Wi-Fi, InterceptFuzz also performs the best, covering a median of 867 branches beyond the benign trace, against 755 for Concat and 233 for AFLNet, and both differences are statistically significant ($p = 0.011$ and $p < 0.001$). The same holds for lines and functions ($p \le 0.019$). The course of the campaigns, shown in Fig.~\ref{fig:covtime}b, resembles that on OpenThread Nexus. InterceptFuzz overtakes Concat after 2~hours, at the end of the second epoch, and stays ahead from then on, while AFLNet again remains far below both fuzzers.

	On ns-3 LTE, all three fuzzers cover between 525 and 614 branches beyond the benign trace, and none of the differences in branch coverage is statistically significant ($p \ge 0.22$). The likely explanation is the limited scope of the LTE model of ns-3, which implements only a small part of the protocol: it lacks the NAS logic, and its RRC implementation supports only a subset of the RRC messages. The code that the fuzzed messages can reach is therefore small, and all three fuzzers appear to largely exhaust it within the campaign. Fig.~\ref{fig:covtime}c shows that they get there at different paces. Concat reaches nearly its final coverage within the first 30~minutes and gains only about ten branches afterwards. InterceptFuzz overtakes it after about 2~hours, at the end of the second epoch, and changes little in the later epochs. AFLNet starts lowest but keeps gaining branches, overtakes both fuzzers after about 7~hours, and ends with the highest median.
	
	\begin{table}[t]
		\centering
		\caption{Distinct crash signatures in OpenThread Nexus over 10$\times$24\,h campaigns: number of campaigns that exposed each signature and median time to first exposure (TTE) over those campaigns. AFLNet exposed none. IF: InterceptFuzz (five epochs, linear schedule). SEGV: segmentation fault. SBO: stack buffer overflow. R/W: read/write.}
		\label{tab:bugs}
		\setlength{\tabcolsep}{3pt}
		\begin{tabular}{lllcccc}
			\hline
			& & & \multicolumn{2}{c}{Campaigns} & \multicolumn{2}{c}{Median TTE (h)} \\
			\cline{4-7}
			ID & Function & Type & Concat & IF & Concat & IF \\
			\hline
			C1 & \texttt{Remove} & Assertion & 9/10 & \textbf{10/10} & 4.1 & \textbf{3.7} \\
			C2 & \texttt{PrepareMeshFrame} & Assertion & 10/10 & 10/10 & \textbf{0.5} & 1.7 \\
			C3 & \texttt{IsRunning} & SEGV (R) & 6/10 & \textbf{9/10} & 8.6 & \textbf{4.2} \\
			C4 & \texttt{SetRloc} & SBO (W) & 7/10 & 7/10 & \textbf{8.9} & 10.9 \\
			C5 & \texttt{IsStable} & SBO (R) & 1/10 & \textbf{3/10} & 10.2 & \textbf{9.9} \\
			C6 & \texttt{IsDhcp} & SBO (R) & 7/10 & 7/10 & 7.8 & \textbf{4.9} \\
			C7 & \texttt{GetNext} & SBO (R) & 4/10 & 4/10 & 6.6 & \textbf{5.3} \\
			\hline
			\multicolumn{3}{l}{Signatures per campaign (median)} & 5 & 5 & & \\
			\hline
		\end{tabular}
	\end{table}
	
	\rqanswer{1}{InterceptFuzz explores more of the targets than both baselines on two of the three targets. On OpenThread Nexus and ns-3 Wi-Fi, it covers significantly more branches beyond the benign trace than Concat and AFLNet. On ns-3 LTE, whose reachable code is small, all three fuzzers reach similar coverage.}

	\subsection{RQ2: Bug Finding}
	\label{sec:eval:bugs}
	RQ2 compares the crash signatures that the three fuzzers expose on each target, deduplicated as described in Section~\ref{sec:eval:setup}. It has two parts, each answered by its own measure. The number of distinct signatures, across all campaigns and per campaign, shows whether a fuzzer exposes more crashes. The number of campaigns in which each signature appears shows how reliably it exposes them. Table~\ref{tab:bugs} reports this number for each signature in OpenThread Nexus, together with the median time to first exposure as additional information. Table~\ref{tab:ns3_crashes} reports the number of distinct signatures for the ns-3 targets. The results are presented per target.

	\subsubsection{OpenThread Nexus}
	Table~\ref{tab:bugs} lists the seven distinct crash signatures found in OpenThread Nexus. All signatures except C2 lie in the handling of Thread Network Data. C1 and C2 are assertion failures, raised when Network Data is removed and when a message is framed for transmission. C3 is an invalid read in the timer code, triggered when the Leader processes a Network Data registration. C4--C7 are stack buffer overflows with a shared call path and likely a shared root cause: when a node appends its Network Data to an MLE message, it walks malformed TLVs past the end of a stack buffer, and the type of the first TLV accessed past the end determines the signature. All seven signatures lie in the OpenThread core (\texttt{src/core}) rather than in the Nexus platform, so they are not artifacts of the simulation and they might be reachable by the attacker described in Section~\ref{sec:eval:attacker}. This also applies to the assertion failures C1 and C2. In builds with assertions enabled, a failed assertion in OpenThread halts the node, which enables a denial of service.

	Both Concat and InterceptFuzz expose the same seven signatures. InterceptFuzz exposes every signature in at least as many campaigns as Concat, and the gap is largest for the rarer signatures: C3 is found in 9 instead of 6 campaigns and C5 in 3 instead of 1. In total, InterceptFuzz exposes the seven signatures 50 times over its ten campaigns, against 44 times for Concat. Per campaign, this difference is too small to be statistically significant: both fuzzers expose a median of five distinct signatures (2--7 for InterceptFuzz and 1--7 for Concat, $p = 0.59$). The median time to first exposure is lower for InterceptFuzz on five of the seven signatures. Note, the median times are computed only over the campaigns that exposed a signature and should therefore be read together with the campaign counts.
	
	AFLNet exposes none of the seven signatures. The final coverage shows that AFLNet does not reach the code in which the signatures lie. In none of its campaigns does it execute a TMF handler, the processing of Network Data registrations by the Leader, or the function in which C2 lies. AFLNet thus remains within the MLE attach, whereas all signatures except C2 lie in the handling of Network Data that follows it. The state model of AFLNet does contain one TMF message type, an Address Query, but the handler of this message is not executed in any of its campaigns either. Why AFLNet does not progress beyond the attach on this target remains unclear and is left for future work.
	\subsubsection{ns-3}
	\begin{table}[t]
		\centering
		\caption{Distinct crash signatures exposed in the ns-3 targets (10 campaigns per fuzzer): union over all campaigns and median [min--max] per campaign. InterceptFuzz uses five epochs under the linear schedule. All signatures are assertion failures in ns-3.}
		\label{tab:ns3_crashes}
		\begin{tabular}{lcc}
			\hline
			\textbf{Fuzzer} & \textbf{All campaigns} & \textbf{Per campaign} \\
			\hline
			\multicolumn{3}{l}{\textit{ns-3 Wi-Fi}} \\
			\quad AFLNet & \textbf{21} & 14.5 [7--18] \\
			\quad Concat & 17 & 13 [11--15] \\
			\quad InterceptFuzz & 20 & \textbf{15} [13--19] \\
			\multicolumn{3}{l}{\textit{ns-3 LTE}} \\
			\quad AFLNet & \textbf{11} & 8.5 [6--10] \\
			\quad Concat & 10 & 8.5 [7--9] \\
			\quad InterceptFuzz & \textbf{11} & \textbf{9} [8--10] \\
			\hline
		\end{tabular}
	\end{table}
	
	In the ns-3 targets, all crash signatures are assertion failures, most of them raised during the deserialization of frames and their information elements. Unlike in OpenThread, these crashes are not vulnerabilities, because, as discussed in the attacker model of Section~\ref{sec:eval:attacker}, ns-3 is not deployed in production. They are nevertheless counted, with the same deduplication as above, as a measure of how many distinct failure sites each fuzzer reaches. Table~\ref{tab:ns3_crashes} summarizes the results for both ns-3 targets.
	
	On ns-3 Wi-Fi, AFLNet exposes 21 distinct signatures across all campaigns, InterceptFuzz 20, and Concat 17. AFLNet thus exposes the most signatures on this target, even though it reaches the lowest coverage of the three fuzzers. Per campaign, InterceptFuzz exposes significantly more signatures than Concat ($p = 0.005$, $\hat{A}_{12} = 0.86$), whereas the difference to AFLNet is not statistically significant ($p = 0.17$).
	
	On ns-3 LTE, InterceptFuzz and AFLNet each expose 11 distinct signatures across all campaigns, against 10 for Concat. Per campaign, InterceptFuzz exposes a median of 9 signatures against 8.5 for both baselines, and neither difference is statistically significant ($p = 0.085$ and $p = 0.13$).
	
	On both ns-3 targets, InterceptFuzz thus exposes at least as many distinct crash signatures per campaign as either baseline, and significantly more than Concat on ns-3 Wi-Fi. Unlike on OpenThread Nexus, InterceptFuzz and AFLNet expose a comparable number of crash signatures on both ns-3 targets. A likely reason is that most of these signatures are raised during the deserialization of frames, which every received frame passes through. Exposing them therefore depends less on reaching deep states than on mutating the frame contents.
	
	\rqanswer{2}{Partly. On OpenThread Nexus, InterceptFuzz and Concat both expose the same seven signatures, while AFLNet exposes none. InterceptFuzz exposes each signature in at least as many campaigns as Concat, and the rarer ones in clearly more. On the ns-3 targets, its advantage is smaller. It exposes significantly more signatures per campaign than Concat only on ns-3 Wi-Fi, and AFLNet exposes about as many signatures as InterceptFuzz on both targets, with one more across all campaigns on ns-3 Wi-Fi.}

	\subsection{RQ3: Prefix Tree Dynamics}
	\label{sec:eval:tree}
	The Prefix Tree is the state model of InterceptFuzz: in every epoch, Phase~I adds new states to it, Phase~II fuzzes the target from these states, and Phase~III removes redundant ones. If Phase~III did not keep the growth of the tree in check, the engine would have to divide its time in Phase~II among an ever larger number of seeds, and thus of states. This section therefore examines how the tree evolves over a campaign in the default configuration on OpenThread Nexus.

	\begin{table*}[t]
		\centering
		\caption{Prefix Tree per epoch on OpenThread Nexus (InterceptFuzz, five epochs, linear schedule). Ph.~I: tree nodes (excluding the root) after Phase~I, i.e., the tree that Phase~II fuzzes. Ph.~III: tree nodes after Phase~III, i.e., the tree that Phase~I of the next epoch extends. Rem.: share of nodes removed by Phase~III. Max.\ depth: depth of the deepest node after Phase~III. Branches: branches covered at the end of the epoch. Medians over ten campaigns, with min--max in brackets.}
		\label{tab:tree}
		\setlength{\tabcolsep}{6pt}
		\begin{tabular}{crrrrcr}
			\hline
			& & \multicolumn{3}{c}{\textbf{Tree nodes}} & & \\
			\textbf{Epoch} & \textbf{End (h)} & \textbf{Ph.~I} & \textbf{Ph.~III} & \textbf{Rem.} & \textbf{Max.\ depth} & \textbf{Branches} \\
			\hline
			1 & 1.6 & 4 {\scriptsize [4--4]} & 3 {\scriptsize [3--3]} & 25\% {\scriptsize [25--25]} & 3 {\scriptsize [3--3]} & 5,194 {\scriptsize [5,002--5,284]} \\
			2 & 4.8 & 2,071 {\scriptsize [1,844--2,408]} & 483 {\scriptsize [400--661]} & 77\% {\scriptsize [73--79]} & 11 {\scriptsize [8--28]} & 6,191 {\scriptsize [5,975--6,405]} \\
			3 & 9.6 & 2,748 {\scriptsize [2,554--3,094]} & 966 {\scriptsize [815--1,189]} & 65\% {\scriptsize [62--68]} & 20 {\scriptsize [15--39]} & 6,599 {\scriptsize [6,375--6,728]} \\
			4 & 16.0 & 2,952 {\scriptsize [2,680--3,339]} & 1,339 {\scriptsize [1,059--1,632]} & 56\% {\scriptsize [51--60]} & 28 {\scriptsize [17--47]} & 6,780 {\scriptsize [6,604--6,990]} \\
			5 & 24.0 & 3,084 {\scriptsize [2,700--3,433]} & 1,565 {\scriptsize [1,337--1,845]} & 49\% {\scriptsize [45--52]} & 46 {\scriptsize [18--67]} & 6,915 {\scriptsize [6,715--7,116]} \\
			\hline
		\end{tabular}
	\end{table*}
	
	Table~\ref{tab:tree} shows the size and depth of the tree after each epoch. In the first epoch, the tree consists of the four packets of the benign trace. Phase~III already removes the last of them, as no transitions can follow the final packet. After the first fuzzing epoch, Phase~I extends the tree by about 2,100 nodes, of which Phase~III keeps about a quarter. In the following epochs, the tree fuzzed in Phase~II levels off at about 3,000 nodes, while the part that survives minimization keeps growing, from 483 to 1,565 nodes, and Phase~III removes a decreasing share of the tree, from 77\% to 49\%. Phase~III thus keeps the growth of the tree in check: summed over the five epochs, Phase~I adds a median of about 8,300 nodes per campaign, whereas the tree after the last Phase~I comprises a median of 3,084 nodes. At the same time, minimization does not discard the progress of earlier epochs, as the retained part of the tree grows in every epoch. The tree also grows deeper: the deepest state retained after the last epoch lies a median of 46 packets after the start of the procedure, compared with 4 packets in the benign trace. InterceptFuzz therefore reaches and keeps states far beyond the benign trace, and keeps them reachable through the stored prefixes rather than through the fuzzer's own mutations. On the two ns-3 targets, the tree evolves in the same way at a smaller scale, ending with a median of 370 nodes after the last Phase~I on ns-3 Wi-Fi and 29 on ns-3 LTE. Appendix~\ref{app:tree} lists the per-epoch values of both targets.

	The last column of Table~\ref{tab:tree} relates this growth to branch coverage. In the first epoch, which fuzzes only the states of the benign trace, InterceptFuzz covers 5,194 branches, fewer than the 5,481 branches that Concat covers in the same 1.6~hours. After the first tree extension, the second epoch adds 997 branches in 3.2~hours, and from then on InterceptFuzz stays ahead of Concat, with 6,191 against 5,979 branches after 4.8~hours. Every InterceptFuzz campaign reaches the final median coverage of Concat (6,603 branches after 24~hours), after a median of 10.5~hours [7.3--16.0]. In the later epochs, in which the fuzzed tree grows only from 2,748 to 3,084 nodes, the gains shrink to 408, 181, and 135 branches, although the epochs become longer. Diminishing gains are typical of any fuzzing campaign, so these numbers alone do not show how much of the slowdown is due to the slower tree growth. The comparison with Concat in Fig.~\ref{fig:covtime}a, however, shows that InterceptFuzz overtakes the baseline only once the tree has been extended beyond the benign trace.
	
	\rqanswer{3}{On OpenThread Nexus, the Prefix Tree grows quickly once the first fuzzing epoch has ended and then levels off. Phase~III removes a large share of the nodes in every epoch and thus keeps the growth of the tree in check, without discarding the progress of earlier epochs, as the retained tree grows in size and depth in every epoch. On the two ns-3 targets, the tree follows the same pattern at a smaller scale.}

	\subsection{RQ4: Epoch Settings}
	\label{sec:eval:schedules}
	\begin{table}[t]
		\centering
		\caption{Epoch settings on OpenThread Nexus (10 campaigns each): final branch coverage, median [min--max], compared with the default 5L (two-sided Mann-Whitney $U$ test and $\hat{A}_{12}$, the probability that a 5L campaign covers more branches), and distinct crash signatures per campaign, median [min--max], and over all campaigns. L: linear schedule. U: uniform schedule.}
		\label{tab:schedules}
		\setlength{\tabcolsep}{3pt}
		\begin{tabular}{lrccrc}
			\hline
			& & \multicolumn{2}{c}{\textbf{5L vs.\ setting}} & \multicolumn{2}{c}{\textbf{Signatures}} \\
			\textbf{Setting} & \textbf{Branches} & $p$ & $\hat{A}_{12}$ & \textbf{Per camp.} & \textbf{All} \\
			\hline
			5L (default) & \textbf{6,915} [6,715--7,116] & & & \textbf{5} [2--7] & \textbf{7} \\
			5U & 6,818 [6,723--6,976] & 0.186 & 0.68 & \textbf{5} [1--6] & \textbf{7} \\
			10L & 6,737 [6,539--6,934] & 0.017 & 0.82 & 4.5 [1--7] & \textbf{7} \\
			10U & 6,883 [6,646--7,145] & 0.473 & 0.60 & \textbf{5} [2--6] & 6 \\
			\hline
		\end{tabular}
	\end{table}

	For OpenThread Nexus, Table~\ref{tab:schedules} compares the default setting, five epochs under the linear schedule (5L), with five epochs under the uniform schedule (5U) and ten epochs under the linear (10L) and under the uniform schedule (10U). The default (5L) reaches the highest median branch coverage, but only the difference to 10L is statistically significant ($p = 0.017$, $\hat{A}_{12} = 0.82$). For 5U ($p = 0.19$, $\hat{A}_{12} = 0.68$) and 10U ($p = 0.47$), ten campaigns are not sufficient to establish a difference. The number of crash signatures per campaign differs little between the settings, with medians of 4.5 to 5, and no difference to the default is statistically significant. Appendix~\ref{app:tree} lists the per-epoch values of all four settings. All four settings end with Prefix Trees of similar size: after the last Phase~I, the medians range from 2,844 to 3,084 nodes, and after the last Phase~III from 1,366 to 1,694 nodes.

	The coverage at the end of each epoch shows where the settings diverge. Under 5U, the tree is extended for the first time only after 4.8~hours. At that point, 5U covers a median of 5,479 branches, whereas 5L and 10U, whose trees have already been extended after 1.6 and 2.4~hours, cover 6,191 and 6,113 branches. 5U narrows this gap in the later epochs but does not close it. This supports the rationale of the linear schedule: the earlier the tree grows beyond the benign trace, the earlier the fuzzer reaches the states that the trace does not contain.
	
	10L extends the tree after only 26~minutes and leads 5L during the first hours, but falls behind between 6.5 and 9.2~hours and never catches up. A plausible reason is that its short first epoch leaves the states of the benign trace under-explored. After the first tree extension, these states have a median of 160 child nodes under 10L, against 229 to 277 under the other settings, and this gap persists until the end of the campaign. The number of child nodes is not the only factor, as 5U has the most children but not the highest coverage, and confirming the mechanism would require recording how much fuzzing time each state receives. The length of the first epoch thus appears to be a trade-off: a long first epoch delays the first tree extension (5U), whereas a very short one leaves the states of the benign trace under-explored (10L).
	
	\rqanswer{4}{The experiments on OpenThread Nexus show that InterceptFuzz is only moderately sensitive to the epoch settings. Of the three alternatives to the default, only ten epochs under the linear schedule reach significantly lower coverage. The length of the first epoch appears to have a larger impact than the number of epochs or the schedule as such.}

	\subsection{RQ5: Integration Effort}
	\label{sec:eval:effort}
	As Table~\ref{tab:targets} shows, integrating InterceptFuzz required 11 hook calls in ns-3 LTE and a single hook call each in ns-3 Wi-Fi and OpenThread Nexus. This difference reflects how the transmit paths of the targets are structured rather than the protocols themselves. ns-3 LTE serializes every RRC message in a dedicated send function, whereas ns-3 Wi-Fi passes all frames through one MAC queue and OpenThread Nexus through one simulated radio. Line counts include only code added or modified for fuzzing, excluding comments, debugging output, build files, and changes that any fuzzing campaign of the target needs, such as disabling logging or encryption. Changes to the target and to the fuzzing harness are reported separately.
	
	The changes to the target are small. In ns-3 LTE, an 8-line helper copies the serialized RRC message, passes it to the hook, and wraps the result in a new packet. Each of the 11 send functions then calls this helper, for 21 lines in total. In ns-3 Wi-Fi, the hook in the MAC queue mutates the frame body and rebuilds the frame with the original MAC header, in 11 lines. In OpenThread Nexus, the hook in the simulated radio takes 6 lines, including the selection of the nodes whose packets are fuzzed.
	
	The harnesses for ns-3 are short new programs of 59 and 48 lines, built from the standard ns-3 helpers. They set up the two nodes, seed ns-3's random number generator, and run the scenario from the entry point. In OpenThread Nexus, the harness reuses the existing integration test \texttt{test\_form\_join.cpp}: 19 lines were added and 47 removed. The changes drop the nodes not involved in the scenario together with logging and assertions, and replace \texttt{main()} with the \texttt{INTERCEPTFUZZ\_TEST} macro, which provides the libFuzzer-style entry point. Any other simulation test can be turned into a harness in the same way.
	
	\rqanswer{5}{Integrating InterceptFuzz requires little manual effort. The changes to the targets range from 6 to 21 lines, and the harnesses are either short programs built from standard helpers or an existing integration test adapted with a few dozen changed lines.}

	\section{Discussion}
	\label{sec:discussion}
	The evaluation shows that InterceptFuzz enables stateful fuzzing of single-process targets without modifying the underlying engine and with limited integration effort, and that Phase~III keeps the growth of its Prefix Tree in check over long campaigns. This section first discusses how to configure InterceptFuzz in practice and which design choices remain open, and then turns to its limitations and to threats to the validity of the evaluation. It closes with further applications of InterceptFuzz.

	\subsection{Choosing the Epoch Settings}
	Section~\ref{sec:eval:schedules} shows that the length of the first epoch matters most: it should be long enough to explore the states of the benign trace, but not so long that it delays the first extension of the Prefix Tree. On the two ns-3 targets, five epochs under the uniform and under the linear schedule perform alike. Five epochs under the linear schedule performed best on OpenThread Nexus. However, the best setting may differ for other targets.

	\subsection{Protecting the State~ID}
	As described in Section~\ref{sec:phase2}, InterceptFuzz deliberately leaves the State~ID exposed to mutation, so that a mutated but valid identifier delivers a packet in a different state. A custom mutator could instead protect the identifier and leave the choice of the state entirely to the seed scheduling of the engine. In preliminary experiments, such a variant achieved slightly lower coverage than the default design and was therefore not pursued further. A systematic comparison of both designs is left for future work.

	\subsection{Concatenation as a Baseline}
	Evaluations of stateful fuzzers commonly include AFLnwe~\cite{aflnwe} as a baseline, for example in ProFuzzBench~\cite{profuzzbench} and in the evaluation of Nyx-Net~\cite{nyxnet}. However, AFLnwe is not only stateless but also unaware of message boundaries. It writes the entire input to the target in a single transmission, so whether the target processes more than one message depends on how it reads its input. LightFTP~\cite{lightftp}, one of the ProFuzzBench targets, for example, treats all bytes received at once as a single command. In contrast, Concat separates the messages by an explicit delimiter and delivers each message separately, as stateful fuzzers do. To the best of the authors' knowledge, such a delimiter-based concatenation is rarely part of the evaluation of stateful fuzzers, although it would arguably provide a stronger lower bound than AFLnwe. In this evaluation, Concat covers more branches than AFLNet in every campaign on OpenThread Nexus and ns-3 Wi-Fi and exposes the same seven crash signatures in OpenThread as InterceptFuzz. For socket-based targets, such a baseline can be built by keeping the message delivery of AFLNet or another socket-based stateful fuzzer and replacing its mutation algorithm with that of Concat.

	\subsection{Relation to ThreadFuzzer}
	ThreadFuzzer~\cite{11659538} also fuzzes OpenThread, but it pursues a different goal. It is a dedicated fuzzer for the Thread protocol that primarily targets physical Thread devices, whereas InterceptFuzz is a generic middleware for stateful fuzzing of single-process targets on top of an unmodified stateless engine. Both were applied to OpenThread, but in different settings. ThreadFuzzer runs every node as a separate process of the OpenThread network simulator, which mirrors its setup for physical devices, whereas InterceptFuzz uses the single-process Nexus framework. The two also intercept packets at different layers. ThreadFuzzer mutates the fields of MLE messages while the protocol stack constructs them, whereas the hook of InterceptFuzz intercepts complete IEEE~802.15.4 frames, so that every packet exchanged between the nodes can be mutated, including TMF messages. Their fuzzing approaches differ as well. ThreadFuzzer mutates benign packets, either randomly or with per-field mutation probabilities adapted from coverage feedback, without keeping a corpus or a model of the states it reaches. InterceptFuzz instead retains the inputs that increase coverage and delivers them in recorded states. Finally, the two tools were applied to different versions of OpenThread, so their results are not compared directly.

	\subsection{Fuzzing Throughput}
	\label{sec:disc:throughput}
	In all three targets, the cost of an execution is dominated by the simulation of the scenario rather than by the fuzzer itself, and it therefore depends on how the mutated packets change the course of the exchange. The targets differ in how a disrupted exchange affects this cost. OpenThread Nexus simulates a fixed 18~seconds, in which a successful attach is followed by further traffic between the nodes, so an input that breaks the attach early leaves fewer packets to simulate. ns-3 LTE instead ends the run as soon as the RRC connection is established, so an input that breaks the attach keeps the run going until the timeout. ns-3 Wi-Fi always simulates a fixed 300~ms.

	The throughput in Table~\ref{tab:coverage} is consistent with this difference. Each input of Concat replaces every packet of the exchange that it covers, whereas InterceptFuzz mutates a single packet after an unmodified prefix, so Concat likely breaks the exchange more often. On OpenThread Nexus, this makes its executions cheaper, and it executes 1.3 times as many inputs per second as InterceptFuzz. On ns-3 LTE, it makes them more expensive, and InterceptFuzz executes 2.4 times as many. On ns-3 Wi-Fi, whose simulated time is fixed, the two differ only slightly. The throughput of AFLNet, in turn, may also be affected by its different underlying engine, LibAFL.

	The throughput of a fuzzer thus depends not only on the fuzzer but also on how the target reacts to a broken exchange. Concat gains throughput where a broken exchange ends the simulated work early, which partly explains why it is a strong baseline on OpenThread Nexus and a weaker one on the ns-3 targets.

	\subsection{Limitations}
	\textbf{Determinism.} Like AFLNet~\cite{meng2025aflnet} and LibAFLstar~\cite{LibAFLStar}, which also reach states by replaying message sequences, InterceptFuzz assumes that the target behaves deterministically, so that the prefix of a State~ID always leads to the same state. As described in Section~\ref{sec:eval:setup}, the evaluated harnesses ensure this by fixing all random seeds. Targets whose behavior depends on wall-clock time or thread interleavings, as well as physical devices, fall outside this model. InterceptFuzz does not check whether a replayed prefix triggers the hook calls in the recorded order, so a divergence would go unnoticed and deliver the payload in a different state than intended. How often this occurs was not measured in the evaluation.

	\textbf{Cost of deep states.} Like the concatenation baseline, every execution starts from the initial state and replays the full prefix of the addressed state, so executions at deep states are more expensive. Combining InterceptFuzz with snapshot-based techniques~\cite{nyxnet,snapfuzz}, which restore a saved state instead of replaying the prefix, could remove this cost and is left for future work.

	\textbf{One freshly mutated packet per execution.} Each execution mutates a single packet after the replayed prefix. Sequences with several mutated packets arise only across epochs, as Phase~I adds the mutated packets of retained inputs to the Prefix Tree, so that a sequence fuzzed in epoch~$N$ contains at most $N$ mutated packets. Bugs that require several packets to be mutated together may therefore take longer to reach than with Concat, which can mutate every packet of a sequence in a single execution.

	\textbf{Scope of the scenario.} InterceptFuzz explores the protocol behavior that the scenario of the harness can lead to, so procedures that the scenario never triggers remain out of reach. The evaluated ns-3 scenarios, for example, contain a single base station and a single access point, so handover in LTE and roaming between access points in Wi-Fi cannot occur. Covering such procedures requires further scenarios, each with its own harness.

	\textbf{Equivalent states.} The Prefix Tree identifies a state by the packet sequence that leads to it. Different prefixes may lead to the same internal state, but determining this exactly would require comparing the complete program state, and even then, irrelevant differences such as counters or memory addresses would obscure equivalent states. InterceptFuzz does not attempt to detect equivalent states and instead keeps its design simple, with every prefix forming a node of its own.

	\textbf{Deployment in OSS-Fuzz.} InterceptFuzz is designed to run on the engines that OSS-Fuzz deploys, but it has not been deployed there. An input \texttt{[State~ID][payload]} reproduces a crash only together with the Prefix Tree that defines its State~ID. For local use, InterceptFuzz provides a script that converts such inputs into self-contained inputs that encode the full packet sequence and are replayed by a dedicated harness. Running InterceptFuzz in a continuous fuzzing service would additionally require keeping the Prefix Tree consistent with a corpus that the service manages on its own, for example when it prunes the corpus. This integration is left for future work.

	\textbf{Manual integration.} Integrating InterceptFuzz into a target requires placing the hook calls by hand and writing a fuzzing harness that runs the scenario. This effort is limited, however: as Section~\ref{sec:eval:effort} shows, the evaluated targets required at most 11 hook calls and 21 changed lines, all at points where packets pass from one node to another. If the target already provides an integration test of the scenario, the harness can be derived from it, as for OpenThread Nexus. A developer familiar with the code base can carry out both steps quickly, and large language models could likely assist with them.

	\subsection{Threats to Validity}
	With ten campaigns per configuration, small differences may remain undetected, and some results lie close to the significance threshold. The AFLNet baseline is the LibAFL reimplementation released with Fizzle, adapted to the interception points of InterceptFuzz. As AFLNet infers its state model from response codes, which the evaluated protocols do not provide, applying it to these targets requires mapping each captured response to a target-specific code. Without such a mapping, AFLNet cannot be applied to these targets at all. A different mapping, or the original implementation of AFLNet, could nevertheless lead to different results. Finally, the evaluation covers three targets built on two simulators. They show that InterceptFuzz applies to different protocols, but not that its results carry over to every single-process target. The targets were hand-picked by the authors for this evaluation, as no established benchmark covers this setting. ProFuzzBench~\cite{profuzzbench}, the common benchmark for stateful fuzzers, consists of network servers that run as separate processes and communicate over sockets, and none of its targets implements a wireless protocol.

	\subsection{Applicability Beyond Protocol Simulators}
	To apply InterceptFuzz to a target, it suffices that all communicating parties run deterministically within a single process and that their messages pass a point at which they can be intercepted. Beyond protocol simulators, this also holds for deterministic simulation testing of distributed systems such as FoundationDB, which simulates an entire cluster within a single-threaded process~\cite{foundationdb}. Since InterceptFuzz exposes a libFuzzer-style entry point, other stateless engines, such as libFuzzer itself, can in principle be used in place of AFL++.
	
	\section{Conclusion}
	\label{sec:conclusion}
	Testing wireless protocol stacks efficiently is difficult, as physical devices are slow, costly to scale, and hard to observe. Many stacks therefore provide simulation platforms that replace the radio with a software model. Single-process simulations, in which all nodes run within one process, go one step further. They offer high throughput and deterministic execution, which makes them well suited to fuzzing, but they contain no sockets for existing stateful fuzzers to attach to.
	
	This paper presented InterceptFuzz, which intercepts packets inside such simulations and lets an unmodified stateless fuzzing engine fuzz them in different protocol states. Each seed carries a State~ID, which selects the state in which its packet is delivered. As the engine remains unmodified, InterceptFuzz inherits its mutators, scheduling, and instrumentation, as well as their future improvements, without reimplementing any of them for stateful fuzzing. The states are recorded in a Prefix Tree, from which InterceptFuzz reaches each state by deterministically replaying the packets that lead to it. The tree grows over the campaign as fuzzing uncovers new states.

	Integrating InterceptFuzz into a target requires only a hook where packets pass between nodes and a harness that runs the scenario, which on the evaluated targets amounted to at most 21 changed lines of target code.
	InterceptFuzz was evaluated on three targets, OpenThread Nexus and the LTE and Wi-Fi models of ns-3, against the delimiter-based concatenation baseline Concat and an interception-based adaptation of AFLNet. On OpenThread Nexus and ns-3 Wi-Fi, it covers significantly more branches than both baselines. On ns-3 LTE, all three fuzzers perform similarly, likely because the fuzzed scenario contains little code beyond the connection setup. During the experiments, seven previously unknown crash signatures were found in OpenThread, which could affect commercially deployed Thread devices. The experiments also showed that InterceptFuzz is only moderately sensitive to its epoch settings, of which the length of the first epoch matters most.

	Future work includes evaluating InterceptFuzz on further single-process targets that, like OpenThread, implement complex protocols with many message types. As Section~\ref{sec:bg:simulation} discusses, such targets are currently rare among open-source wireless stacks. Should more stacks adopt single-process simulation, a benchmark of such targets would allow stateful fuzzers to be compared in this setting, with InterceptFuzz and Concat as reference points.

	\section*{Acknowledgment}
	This work was supported by the Flemish Government through the Cybersecurity Research Program with grant number VOEWICS02.

	\bibliographystyle{ieeetr}
	\bibliography{references,references_extra}

\begin{thebibliography}{10}

\bibitem{OSS_Fuzz}
Google, ``{OSS-Fuzz - continuous fuzzing for open source software}.''
  \url{https://github.com/google/oss-fuzz}, 2026.
\newblock [Accessed 30-Oct-2026].

\bibitem{libfuzzer}
{LLVM Project}, ``{libFuzzer – a library for coverage-guided fuzz testing}.''
  \url{https://llvm.org/docs/LibFuzzer.html}.
\newblock [Accessed 30-Oct-2026].

\bibitem{aflplusplus}
A.~Fioraldi, D.~Maier, H.~Ei{\ss}feldt, and M.~Heuse, ``{AFL++}: Combining
  incremental steps of fuzzing research,'' in {\em 14th {USENIX} Workshop on
  Offensive Technologies ({WOOT} 20)}, {USENIX} Association, Aug. 2020.

\bibitem{honggfuzz}
Google, ``{honggfuzz}.'' \url{https://github.com/google/honggfuzz}, 2026.
\newblock [Accessed 30-Oct-2026].

\bibitem{centipede}
Google, ``{Centipede}: A distributed fuzzing engine.''
  \url{https://github.com/google/fuzztest/tree/main/centipede}, 2026.
\newblock [Accessed 30-Oct-2026].

\bibitem{ns3}
G.~F. Riley and T.~R. Henderson, ``The ns-3 network simulator,'' in {\em
  Modeling and Tools for Network Simulation}, pp.~15--34, Springer, 2010.

\bibitem{OpenThread}
Google, ``{OpenThread}.'' \url{https://openthread.io/}, 2026.
\newblock [Accessed 30-Oct-2026].

\bibitem{AFLNet}
{Van{-}Thuan Pham and Marcel B{\"o}hme and Abhik Roychoudhury}, ``{AFLNet: A
  Greybox Fuzzer for Network Protocols},'' in {\em Proceedings of the 13th IEEE
  International Conference on Software Testing, Verification and Validation :
  Testing Tools Track}, 2020.

\bibitem{meng2025aflnet}
R.~Meng, V.-T. Pham, M.~B{\"o}hme, and A.~Roychoudhury, ``Aflnet five years
  later: On coverage-guided protocol fuzzing,'' {\em IEEE Transactions on
  Software Engineering}, vol.~51, no.~4, pp.~960--974, 2025.

\bibitem{libafl}
A.~Fioraldi, D.~C. Maier, D.~Zhang, and D.~Balzarotti, ``{LibAFL}: A framework
  to build modular and reusable fuzzers,'' in {\em Proceedings of the 2022 ACM
  SIGSAC Conference on Computer and Communications Security (CCS)},
  pp.~1051--1065, ACM, 2022.

\bibitem{fizzle}
N.~Bennett, T.~Tucker, C.~Stillman, W.~Enck, P.~Traynor, and K.~Butler,
  ``{Fizzle}: A framework for deterministic and reproducible network fuzzing,''
  in {\em 2026 IEEE Symposium on Security and Privacy (SP)}, 2026.

\bibitem{Miller_Fuzzing}
{Miller, Barton P. and Fredriksen, Lars and So, Bryan}, ``{An empirical study
  of the reliability of UNIX utilities},'' {\em Commun. ACM}, vol.~33,
  p.~32–44, Dec. 1990.

\bibitem{daniele2024statefulfuzzingreallychallenging}
C.~Daniele, ``Is stateful fuzzing really challenging?,'' 2024.

\bibitem{stateafl}
R.~Natella, ``{StateAFL: Greybox Fuzzing for Stateful Network Servers},'' {\em
  Empirical Software Engineering}, vol.~27, no.~191, 2022.

\bibitem{sgfuzz}
J.~Ba, M.~B{\"o}hme, Z.~Mirzamomen, and A.~Roychoudhury, ``{Stateful Greybox
  Fuzzing},'' in {\em 31st USENIX Security Symposium (USENIX Security 22)},
  (Boston, MA), pp.~3255--3272, USENIX Association, Aug. 2022.

\bibitem{chatafl}
R.~Meng, M.~Mirchev, M.~B\"{o}hme, and A.~Roychoudhury, ``{Large Language Model
  guided Protocol Fuzzing},'' in {\em Proceedings of the 31st Annual Network
  and Distributed System Security Symposium (NDSS)}, 2024.

\bibitem{snapfuzz}
A.~Andronidis and C.~Cadar, ``{SnapFuzz}: High-throughput fuzzing of network
  applications,'' in {\em Proceedings of the 31st ACM SIGSOFT International
  Symposium on Software Testing and Analysis (ISSTA)}, pp.~340--351, 2022.

\bibitem{libdesock}
{Fraunhofer FKIE}, ``libdesock: De-socketing for fuzzing.''
  \url{https://github.com/fkie-cad/libdesock}.
\newblock [Accessed 30-Oct-2026].

\bibitem{nyxnet}
S.~Schumilo, C.~Aschermann, A.~Jemmett, A.~Abbasi, and T.~Holz, ``{Nyx-Net}:
  Network fuzzing with incremental snapshots,'' in {\em Proceedings of the
  Seventeenth European Conference on Computer Systems (EuroSys)}, 2022.

\bibitem{LibAFLStar}
C.~Daniele, T.~Bethe, M.~Maugeri, A.~Continella, and E.~Poll, ``{LibAFLstar}:
  Fast and state-aware protocol fuzzing,'' in {\em Computer Security -- ESORICS
  2025} (V.~Nicomette, A.~Benzekri, N.~Boulahia-Cuppens, and J.~Vaidya, eds.),
  vol.~16055 of {\em Lecture Notes in Computer Science}, (Cham), pp.~105--123,
  Springer, 2026.

\bibitem{greyhound}
M.~Garbelini, C.~Wang, and S.~Chattopadhyay, ``{Greyhound: Directed Greybox
  Wi-Fi Fuzzing},'' {\em IEEE Transactions on Dependable and Secure Computing},
  vol.~PP, pp.~1--1, 08 2020.

\bibitem{sweyntooth}
M.~E. Garbelini, C.~Wang, S.~Chattopadhyay, S.~Sun, and E.~Kurniawan,
  ``{SweynTooth: unleashing mayhem over bluetooth low energy},'' in {\em
  Proceedings of the 2020 USENIX Conference on Usenix Annual Technical
  Conference}, USENIX ATC'20, (USA), USENIX Association, 2020.

\bibitem{BLuEMan}
W.-C. Kao, Y.-C. Chen, Y.-S. Lin, Y.-C. Yang, C.-Y. Li, and C.-Y. Huang,
  ``{BLuEMan}: A stateful simulation-based fuzzing framework for open-source
  {RTOS} {Bluetooth} low energy protocol stacks,'' in {\em 34th USENIX Security
  Symposium (USENIX Security 25)}, 2025.

\bibitem{braktooth}
M.~E. Garbelini, V.~Bedi, S.~Chattopadhyay, S.~Sun, and E.~Kurniawan,
  ``{BrakTooth: Causing Havoc on Bluetooth Link Manager via Directed
  Fuzzing},'' in {\em 31st USENIX Security Symposium (USENIX Security 22)},
  (Boston, MA), pp.~1025--1042, USENIX Association, Aug. 2022.

\bibitem{Frankenstein}
J.~Ruge, J.~Classen, F.~Gringoli, and M.~Hollick, ``{Frankenstein: Advanced
  Wireless Fuzzing to Exploit New Bluetooth Escalation Targets},'' in {\em 29th
  USENIX Security Symposium (USENIX Security 20)}, pp.~19--36, USENIX
  Association, Aug. 2020.

\bibitem{5ghoul}
{Matheus E. Garbelini, Zewen Shang, Shijie Luo, Sudipta Chattopadhyay, Sumei
  Sun, Ernest Kurniawan}, ``{ASSET Research Group: 5Ghoul}.''
  \url{https://asset-group.github.io/disclosures/5ghoul/}.
\newblock [Accessed 30-Oct-2026].

\bibitem{covfuzz}
I.~Siro{\v{s}}, D.~Singel{\'e}e, and B.~Preneel, ``{CovFUZZ: Coverage-based
  fuzzer for 4G\&5G protocols},'' in {\em {2025 IEEE 10th European Symposium on
  Security and Privacy (EuroS\&P)}}, pp.~737--754, IEEE, 2025.

\bibitem{breakingsilence}
I.~Siro{\v{s}}, R.~Cavalcanti, D.~Singel{\'e}e, and B.~Preneel, ``Breaking the
  silence: Fuzzing {LTE-M} and {NB-IoT} protocols,'' in {\em Computer Security.
  ESORICS 2025 International Workshops}, vol.~16233 of {\em Lecture Notes in
  Computer Science}, pp.~524--543, Springer, 2026.

\bibitem{11659538}
I.~Siroš, J.~Heirwegh, D.~Singelée, and B.~Preneel, ``Threadfuzzer: A fuzzing
  framework for the thread protocol,'' in {\em 2026 22nd International
  Conference on Distributed Computing in Smart Systems and the Internet of
  Things (DCOSS-IoT)}, pp.~464--472, 2026.

\bibitem{PeriScope}
D.~Song, F.~Hetzelt, D.~Das, C.~Spensky, Y.~Na, S.~Volckaert, G.~Vigna,
  C.~Kruegel, J.-P. Seifert, and M.~Franz, ``{PeriScope}: An effective probing
  and fuzzing framework for the hardware-{OS} boundary,'' in {\em Network and
  Distributed System Security Symposium (NDSS)}, 2019.

\bibitem{OpenThread_Simulation}
Google, ``{OpenThread} simulation.''
  \url{https://openthread.io/platforms/simulation}.
\newblock [Accessed 30-Oct-2026].

\bibitem{srsRAN_ZMQ}
{Software Radio Systems}, ``{srsRAN 4G} with {ZMQ} virtual radios.''
  \url{https://docs.srsran.com/projects/4g/en/latest/app_notes/source/zeromq/source/index.html}.
\newblock [Accessed 30-Oct-2026].

\bibitem{srsRAN_Project_ZMQ}
{Software Radio Systems}, ``{srsRAN} {gNB} with {srsUE}.''
  \url{https://docs.srsran.com/projects/project/en/latest/tutorials/source/srsUE/source/index.html}.
\newblock [Accessed 30-Oct-2026].

\bibitem{open5gs_tests}
{Open5GS}, ``{Open5GS} test framework ({tests/common/application.c}).''
  \url{https://github.com/open5gs/open5gs/blob/main/tests/common/application.c}.
\newblock [Accessed 30-Oct-2026].

\bibitem{zephyr_bsim}
{Zephyr Project}, ``{BabbleSim}.''
  \url{https://docs.zephyrproject.org/latest/develop/test/bsim.html}.
\newblock [Accessed 30-Oct-2026].

\bibitem{hostap_hwsim}
J.~Malinen, ``{hostapd/wpa\_supplicant} testing with {mac80211\_hwsim}.''
  \url{https://w1.fi/cgit/hostap/tree/tests/hwsim/README}.
\newblock [Accessed 12-Oct-2026].

\bibitem{klees2018evaluating}
G.~Klees, A.~Ruef, B.~Cooper, S.~Wei, and M.~Hicks, ``Evaluating fuzz
  testing,'' in {\em Proceedings of the 2018 ACM SIGSAC Conference on Computer
  and Communications Security (CCS)}, pp.~2123--2138, ACM, 2018.

\bibitem{arcuri2011practical}
A.~Arcuri and L.~Briand, ``A practical guide for using statistical tests to
  assess randomized algorithms in software engineering,'' in {\em Proceedings
  of the 33rd International Conference on Software Engineering (ICSE)},
  pp.~1--10, ACM, 2011.

\bibitem{aflnwe}
V.-T. Pham, ``{AFLnwe}: A network-enabled version of {AFL}.''
  \url{https://github.com/aflnet/aflnwe}.
\newblock [Accessed 30-Oct-2026].

\bibitem{profuzzbench}
R.~Natella and V.-T. Pham, ``{ProFuzzBench}: A benchmark for stateful protocol
  fuzzing,'' in {\em Proceedings of the 30th ACM SIGSOFT International
  Symposium on Software Testing and Analysis (ISSTA)}, pp.~662--665, 2021.

\bibitem{lightftp}
hfiref0x, ``{LightFTP}: Small x86-32/x64 {FTP} server.''
  \url{https://github.com/hfiref0x/LightFTP}.
\newblock [Accessed 30-Oct-2026].

\bibitem{foundationdb}
J.~Zhou, M.~Xu, A.~Shraer, B.~Namasivayam, A.~Miller, E.~Tschannen,
  S.~Atherton, A.~J. Beamon, R.~Sears, J.~Leach, D.~Rosenthal, X.~Dong,
  W.~Wilson, B.~Collins, D.~Scherer, A.~Grieser, Y.~Liu, A.~Moore, B.~Muppana,
  X.~Su, and V.~Yadav, ``{FoundationDB}: A distributed unbundled transactional
  key value store,'' in {\em Proceedings of the 2021 International Conference
  on Management of Data (SIGMOD)}, 2021.

\end{thebibliography}

	\appendices
	\section{Prefix Tree Dynamics}
	\label{app:tree}
	Table~\ref{tab:tree_all} extends Table~\ref{tab:tree} to all four epoch settings compared in Section~\ref{sec:eval:schedules}. Table~\ref{tab:tree_ns3} shows the same values for the default setting on the two ns-3 targets.

	\begin{table*}[t]
		\centering
		\caption{Prefix Tree per epoch on OpenThread Nexus for all four epoch settings of InterceptFuzz (10 campaigns each). End (h): end of the epoch's Phase~II within the campaign. Ph.~I: tree nodes (excluding the root) after Phase~I, i.e., the tree that Phase~II fuzzes. Ph.~III: tree nodes after Phase~III, i.e., the tree that Phase~I of the next epoch extends. Rem.: share of nodes removed by Phase~III. Max.\ depth: depth of the deepest node after Phase~III. Branches: branches covered at the end of the epoch. Medians over ten campaigns, with min--max in brackets.}
		\label{tab:tree_all}
		\setlength{\tabcolsep}{6pt}
		\begin{tabular}{crrrrcr}
			\hline
			& & \multicolumn{3}{c}{\textbf{Tree nodes}} & & \\
			\textbf{Epoch} & \textbf{End (h)} & \textbf{Ph.~I} & \textbf{Ph.~III} & \textbf{Rem.} & \textbf{Max.\ depth} & \textbf{Branches} \\
			\hline
			\multicolumn{7}{l}{\textit{5L: five epochs, linear schedule (default)}} \\
			1 & 1.6 & 4 {\scriptsize [4--4]} & 3 {\scriptsize [3--3]} & 25\% {\scriptsize [25--25]} & 3 {\scriptsize [3--3]} & 5,194 {\scriptsize [5,002--5,284]} \\
			2 & 4.8 & 2,071 {\scriptsize [1,844--2,408]} & 483 {\scriptsize [400--661]} & 77\% {\scriptsize [73--79]} & 11 {\scriptsize [8--28]} & 6,191 {\scriptsize [5,975--6,405]} \\
			3 & 9.6 & 2,748 {\scriptsize [2,554--3,094]} & 966 {\scriptsize [815--1,189]} & 65\% {\scriptsize [62--68]} & 20 {\scriptsize [15--39]} & 6,599 {\scriptsize [6,375--6,728]} \\
			4 & 16.0 & 2,952 {\scriptsize [2,680--3,339]} & 1,339 {\scriptsize [1,059--1,632]} & 56\% {\scriptsize [51--60]} & 28 {\scriptsize [17--47]} & 6,780 {\scriptsize [6,604--6,990]} \\
			5 & 24.0 & 3,084 {\scriptsize [2,700--3,433]} & 1,565 {\scriptsize [1,337--1,845]} & 49\% {\scriptsize [45--52]} & 46 {\scriptsize [18--67]} & 6,915 {\scriptsize [6,715--7,116]} \\
			\multicolumn{7}{l}{\textit{5U: five epochs, uniform schedule}} \\
			1 & 4.8 & 4 {\scriptsize [4--4]} & 3 {\scriptsize [3--3]} & 25\% {\scriptsize [25--25]} & 3 {\scriptsize [3--3]} & 5,479 {\scriptsize [5,333--5,569]} \\
			2 & 9.6 & 2,382 {\scriptsize [2,208--2,756]} & 603 {\scriptsize [522--748]} & 75\% {\scriptsize [73--78]} & 15 {\scriptsize [10--34]} & 6,287 {\scriptsize [6,079--6,473]} \\
			3 & 14.4 & 2,927 {\scriptsize [2,619--3,161]} & 1,047 {\scriptsize [881--1,131]} & 64\% {\scriptsize [62--67]} & 20 {\scriptsize [14--37]} & 6,641 {\scriptsize [6,337--6,814]} \\
			4 & 19.2 & 2,928 {\scriptsize [2,523--3,215]} & 1,262 {\scriptsize [1,070--1,526]} & 57\% {\scriptsize [51--59]} & 21 {\scriptsize [17--41]} & 6,750 {\scriptsize [6,507--6,891]} \\
			5 & 24.0 & 2,931 {\scriptsize [2,661--3,454]} & 1,366 {\scriptsize [1,138--1,692]} & 54\% {\scriptsize [47--57]} & 24 {\scriptsize [19--41]} & 6,818 {\scriptsize [6,723--6,976]} \\
			\multicolumn{7}{l}{\textit{10L: ten epochs, linear schedule}} \\
			1 & 0.4 & 4 {\scriptsize [4--4]} & 3 {\scriptsize [3--3]} & 25\% {\scriptsize [25--25]} & 3 {\scriptsize [3--3]} & 4,713 {\scriptsize [4,554--4,912]} \\
			2 & 1.3 & 1,656 {\scriptsize [1,477--1,909]} & 342 {\scriptsize [248--462]} & 80\% {\scriptsize [75--85]} & 9 {\scriptsize [8--36]} & 5,343 {\scriptsize [5,042--5,611]} \\
			3 & 2.6 & 2,034 {\scriptsize [1,830--2,478]} & 615 {\scriptsize [521--806]} & 71\% {\scriptsize [65--73]} & 13 {\scriptsize [10--40]} & 5,766 {\scriptsize [5,547--6,169]} \\
			4 & 4.4 & 2,190 {\scriptsize [1,910--2,519]} & 813 {\scriptsize [645--1,020]} & 62\% {\scriptsize [56--68]} & 21 {\scriptsize [12--39]} & 6,154 {\scriptsize [5,948--6,407]} \\
			5 & 6.5 & 2,441 {\scriptsize [2,153--3,273]} & 1,128 {\scriptsize [928--1,557]} & 53\% {\scriptsize [51--60]} & 24 {\scriptsize [20--39]} & 6,354 {\scriptsize [6,140--6,480]} \\
			6 & 9.2 & 2,493 {\scriptsize [2,250--3,252]} & 1,220 {\scriptsize [1,014--1,676]} & 51\% {\scriptsize [45--58]} & 28 {\scriptsize [22--39]} & 6,468 {\scriptsize [6,218--6,693]} \\
			7 & 12.2 & 2,551 {\scriptsize [2,209--3,342]} & 1,313 {\scriptsize [1,182--1,813]} & 47\% {\scriptsize [44--51]} & 31 {\scriptsize [23--45]} & 6,565 {\scriptsize [6,343--6,827]} \\
			8 & 15.7 & 2,710 {\scriptsize [2,335--3,461]} & 1,491 {\scriptsize [1,261--1,977]} & 45\% {\scriptsize [41--49]} & 32 {\scriptsize [27--49]} & 6,669 {\scriptsize [6,400--6,846]} \\
			9 & 19.6 & 2,757 {\scriptsize [2,383--3,332]} & 1,652 {\scriptsize [1,324--2,042]} & 42\% {\scriptsize [39--45]} & 34 {\scriptsize [27--61]} & 6,691 {\scriptsize [6,485--6,880]} \\
			10 & 24.0 & 2,844 {\scriptsize [2,415--3,447]} & 1,694 {\scriptsize [1,367--2,075]} & 41\% {\scriptsize [35--43]} & 45 {\scriptsize [27--61]} & 6,737 {\scriptsize [6,539--6,934]} \\
			\multicolumn{7}{l}{\textit{10U: ten epochs, uniform schedule}} \\
			1 & 2.4 & 4 {\scriptsize [4--4]} & 3 {\scriptsize [3--3]} & 25\% {\scriptsize [25--25]} & 3 {\scriptsize [3--3]} & 5,235 {\scriptsize [4,984--5,324]} \\
			2 & 4.8 & 2,155 {\scriptsize [2,021--2,426]} & 494 {\scriptsize [442--631]} & 77\% {\scriptsize [74--79]} & 11 {\scriptsize [9--33]} & 6,113 {\scriptsize [5,696--6,223]} \\
			3 & 7.2 & 2,697 {\scriptsize [2,354--2,883]} & 898 {\scriptsize [760--1,066]} & 66\% {\scriptsize [62--70]} & 18 {\scriptsize [12--34]} & 6,260 {\scriptsize [6,075--6,533]} \\
			4 & 9.6 & 2,689 {\scriptsize [2,455--2,916]} & 1,089 {\scriptsize [932--1,223]} & 59\% {\scriptsize [55--64]} & 20 {\scriptsize [17--35]} & 6,363 {\scriptsize [6,236--6,867]} \\
			5 & 12.0 & 2,716 {\scriptsize [2,500--2,930]} & 1,223 {\scriptsize [1,130--1,536]} & 55\% {\scriptsize [48--59]} & 23 {\scriptsize [18--52]} & 6,496 {\scriptsize [6,315--6,997]} \\
			6 & 14.4 & 2,732 {\scriptsize [2,487--3,060]} & 1,309 {\scriptsize [1,177--1,559]} & 52\% {\scriptsize [46--57]} & 24 {\scriptsize [16--51]} & 6,550 {\scriptsize [6,405--7,011]} \\
			7 & 16.8 & 2,813 {\scriptsize [2,494--3,111]} & 1,475 {\scriptsize [1,257--1,648]} & 49\% {\scriptsize [42--54]} & 30 {\scriptsize [16--55]} & 6,699 {\scriptsize [6,437--7,019]} \\
			8 & 19.2 & 2,846 {\scriptsize [2,502--3,127]} & 1,547 {\scriptsize [1,307--1,774]} & 47\% {\scriptsize [40--52]} & 33 {\scriptsize [18--55]} & 6,760 {\scriptsize [6,530--7,022]} \\
			9 & 21.6 & 2,841 {\scriptsize [2,497--3,175]} & 1,575 {\scriptsize [1,328--1,758]} & 45\% {\scriptsize [40--51]} & 34 {\scriptsize [19--55]} & 6,870 {\scriptsize [6,617--7,113]} \\
			10 & 24.0 & 2,864 {\scriptsize [2,498--3,165]} & 1,614 {\scriptsize [1,416--1,828]} & 43\% {\scriptsize [40--52]} & 35 {\scriptsize [19--55]} & 6,883 {\scriptsize [6,646--7,145]} \\
			\hline
		\end{tabular}
	\end{table*}

	\begin{table*}[t]
		\centering
		\caption{Prefix Tree per epoch on the ns-3 targets (InterceptFuzz, five epochs, linear schedule, 10 campaigns each). Columns as in Table~\ref{tab:tree}. Medians over ten campaigns, with min--max in brackets.}
		\label{tab:tree_ns3}
		\setlength{\tabcolsep}{6pt}
		\begin{tabular}{crrrrcr}
			\hline
			& & \multicolumn{3}{c}{\textbf{Tree nodes}} & & \\
			\textbf{Epoch} & \textbf{End (h)} & \textbf{Ph.~I} & \textbf{Ph.~III} & \textbf{Rem.} & \textbf{Max.\ depth} & \textbf{Branches} \\
			\hline
			\multicolumn{7}{l}{\textit{ns-3 Wi-Fi}} \\
			1 & 0.7 & 7 {\scriptsize [7--7]} & 4 {\scriptsize [3--5]} & 43\% {\scriptsize [29--57]} & 4 {\scriptsize [3--5]} & 40,911 {\scriptsize [40,883--41,010]} \\
			2 & 2.0 & 277 {\scriptsize [230--318]} & 77 {\scriptsize [63--89]} & 72\% {\scriptsize [65--77]} & 9 {\scriptsize [7--11]} & 40,994 {\scriptsize [40,969--41,100]} \\
			3 & 4.0 & 303 {\scriptsize [266--345]} & 121 {\scriptsize [104--153]} & 59\% {\scriptsize [55--66]} & 13 {\scriptsize [11--16]} & 41,273 {\scriptsize [41,046--41,328]} \\
			4 & 6.7 & 362 {\scriptsize [300--402]} & 171 {\scriptsize [126--190]} & 53\% {\scriptsize [43--58]} & 17 {\scriptsize [12--19]} & 41,344 {\scriptsize [41,053--41,458]} \\
			5 & 10.0 & 370 {\scriptsize [323--441]} & 201 {\scriptsize [160--254]} & 46\% {\scriptsize [38--53]} & 18 {\scriptsize [15--23]} & 41,363 {\scriptsize [41,054--41,459]} \\
			\multicolumn{7}{l}{\textit{ns-3 LTE}} \\
			1 & 0.7 & 3 {\scriptsize [3--3]} & 2 {\scriptsize [2--2]} & 33\% {\scriptsize [33--33]} & 2 {\scriptsize [2--2]} & 34,821 {\scriptsize [34,809--35,039]} \\
			2 & 2.0 & 34 {\scriptsize [14--44]} & 16 {\scriptsize [6--20]} & 53\% {\scriptsize [53--57]} & 3 {\scriptsize [3--3]} & 34,889 {\scriptsize [34,868--35,091]} \\
			3 & 4.0 & 26 {\scriptsize [17--29]} & 18 {\scriptsize [9--22]} & 33\% {\scriptsize [24--47]} & 5 {\scriptsize [5--5]} & 34,917 {\scriptsize [34,889--35,111]} \\
			4 & 6.7 & 30 {\scriptsize [18--35]} & 21 {\scriptsize [11--26]} & 30\% {\scriptsize [21--41]} & 7 {\scriptsize [5--8]} & 34,919 {\scriptsize [34,889--35,112]} \\
			5 & 10.0 & 29 {\scriptsize [22--36]} & 22 {\scriptsize [15--25]} & 30\% {\scriptsize [18--38]} & 7 {\scriptsize [5--9]} & 34,922 {\scriptsize [34,913--35,113]} \\
			\hline
		\end{tabular}
	\end{table*}

	\section{Coverage Breakdown on OpenThread Nexus}
	\label{app:coverage}
	Table~\ref{tab:cov_breakdown} breaks down the branch coverage of Table~\ref{tab:coverage} on OpenThread Nexus by source directory and lists the files with the largest differences between InterceptFuzz and Concat. The breakdown is computed from gcovr reports of a separate replay of the same inputs as Table~\ref{tab:coverage}. Repeated replays occasionally yield slightly different coverage, so the totals deviate from that table by at most two branches.

	Most additional branches of InterceptFuzz lie in the handling of TMF messages. InterceptFuzz covers a median of 245 CoAP branches against 151 for Concat and 36 for AFLNet, and 365 against 230 and 165 branches in the commissioning and dataset code (\texttt{src/core/meshcop}). Within \texttt{src/core/thread}, the largest gains lie in the Leader's handling of Network Data and in address resolution. Across all campaigns, 771 branches are covered only by InterceptFuzz, against 173 covered only by Concat. In turn, Concat covers more branches in the network layer (\texttt{src/core/net}), mainly because it reaches the DNS-SD server in six of its ten campaigns, against three for InterceptFuzz.

	InterceptFuzz covers a median of 2,592 branches more than AFLNet, two thirds of them in \texttt{src/core/thread}. Most of this difference lies in code that Concat covers as well. AFLNet already covers only about half of the MLE code (\texttt{mle.cpp} and \texttt{mle\_ftd.cpp}) that the other two fuzzers reach, and it covers hardly any of the code that runs after a completed attach, such as the handling of Network Data and the indirect transmission to children. Across all campaigns, 2,999 branches are covered by InterceptFuzz but never by AFLNet, 2,229 of which are also covered by Concat. Only eight branches are covered by AFLNet but never by InterceptFuzz.

	As the benign run ends shortly after the attach, it contains no TMF message. To see whether the Prefix Trees of InterceptFuzz nevertheless reach the states in which TMF messages are exchanged, their packets are classified by decoding their IEEE~802.15.4, 6LoWPAN, and UDP headers: MLE messages use UDP port 19788 and TMF messages port 61631. The type of a TMF request is taken from the resource in its CoAP Uri-Path options. The tree after Phase~I contains no TMF message in the first epoch of any campaign, and a median of 46, 219, 295, and 351 TMF messages in epochs 2 to 5. In the final trees, TMF messages account for 9--18\% of all nodes. The most frequent requests are Address Queries, carried by 622 edges, and Network Data registrations, carried by 508 edges, both occurring in all ten campaigns. They are followed by Address Solicit requests and requests for commissioner and dataset information, each occurring in seven or eight campaigns.

	\begin{table*}[t]
		\centering
		\caption{Branch coverage on OpenThread Nexus per directory of \textnormal{\texttt{src/core}} and for the files with the largest differences between InterceptFuzz (IF) and Concat and between InterceptFuzz and AFLNet, median [min--max] over ten campaigns. IF $-$ Concat and IF $-$ AFLNet: differences of the medians.}
		\label{tab:cov_breakdown}
		\setlength{\tabcolsep}{6pt}
		\begin{tabular}{lrrrrr}
			\hline
			\textbf{Directory / file} & \textbf{AFLNet} & \textbf{Concat} & \textbf{IF} & \textbf{IF $-$ Concat} & \textbf{IF $-$ AFLNet} \\
			\hline
			\texttt{thread} & 2,021 {\scriptsize [1,933--2,087]} & 3,609 {\scriptsize [3,446--3,691]} & 3,759 {\scriptsize [3,669--3,967]} & $+$150 & $+$1,738 \\
			\texttt{meshcop} & 165 {\scriptsize [163--244]} & 230 {\scriptsize [215--325]} & 365 {\scriptsize [314--394]} & $+$135 & $+$200 \\
			\texttt{coap} & 36 {\scriptsize [36--119]} & 151 {\scriptsize [111--217]} & 245 {\scriptsize [243--246]} & $+$94 & $+$209 \\
			\texttt{net} & 527 {\scriptsize [516--668]} & 716 {\scriptsize [561--745]} & 573 {\scriptsize [562--720]} & $-$143 & $+$46 \\
			\texttt{mac} & 499 {\scriptsize [495--505]} & 610 {\scriptsize [560--623]} & 593 {\scriptsize [592--626]} & $-$17 & $+$94 \\
			\texttt{common} & 622 {\scriptsize [608--652]} & 765 {\scriptsize [721--784]} & 758 {\scriptsize [746--780]} & $-$7 & $+$136 \\
			\texttt{utils} & 82 {\scriptsize [81--87]} & 130 {\scriptsize [119--192]} & 141 {\scriptsize [119--197]} & $+$11 & $+$59 \\
			\texttt{crypto} & 110 {\scriptsize [110--110]} & 111 {\scriptsize [111--111]} & 111 {\scriptsize [111--111]} & 0 & $+$1 \\
			Other & 237 {\scriptsize [236--240]} & 300 {\scriptsize [295--315]} & 300 {\scriptsize [296--313]} & 0 & $+$63 \\
			\textbf{Total} & 4,323 {\scriptsize [4,219--4,555]} & 6,602 {\scriptsize [6,250--6,790]} & 6,915 {\scriptsize [6,715--7,116]} & $+$313 & $+$2,592 \\
			\hline
			\multicolumn{6}{l}{\textit{Files with the largest differences between IF and Concat}} \\
			\texttt{coap.cpp} & 18 {\scriptsize [18--48]} & 88 {\scriptsize [54--132]} & 155 {\scriptsize [154--156]} & $+$67 & $+$137 \\
			\texttt{secure\_transport.cpp} & 14 {\scriptsize [14--85]} & 14 {\scriptsize [14--85]} & 79 {\scriptsize [77--85]} & $+$65 & $+$65 \\
			\texttt{address\_resolver.cpp} & 91 {\scriptsize [81--98]} & 128 {\scriptsize [116--134]} & 160 {\scriptsize [137--188]} & $+$32 & $+$69 \\
			\texttt{network\_data\_leader\_ftd.cpp} & 25 {\scriptsize [25--25]} & 134 {\scriptsize [68--143]} & 162 {\scriptsize [132--234]} & $+$28 & $+$137 \\
			\texttt{coap\_message.cpp} & 18 {\scriptsize [18--69]} & 60 {\scriptsize [55--82]} & 86 {\scriptsize [85--86]} & $+$26 & $+$68 \\
			\texttt{dataset\_manager.cpp} & 32 {\scriptsize [32--33]} & 51 {\scriptsize [43--52]} & 74 {\scriptsize [50--77]} & $+$23 & $+$42 \\
			\texttt{tmf.cpp} & 15 {\scriptsize [14--16]} & 19 {\scriptsize [14--21]} & 38 {\scriptsize [33--45]} & $+$19 & $+$23 \\
			\texttt{lowpan.cpp} & 343 {\scriptsize [311--364]} & 367 {\scriptsize [321--382]} & 383 {\scriptsize [360--405]} & $+$16 & $+$40 \\
			\texttt{dns\_types.cpp} & 0 {\scriptsize [0--72]} & 73 {\scriptsize [0--75]} & 0 {\scriptsize [0--75]} & $-$73 & 0 \\
			\texttt{dnssd\_server.cpp} & 5 {\scriptsize [5--65]} & 65 {\scriptsize [5--69]} & 5 {\scriptsize [5--68]} & $-$60 & 0 \\
			\texttt{mesh\_forwarder.cpp} & 212 {\scriptsize [206--218]} & 269 {\scriptsize [257--274]} & 244 {\scriptsize [242--257]} & $-$25 & $+$32 \\
			\texttt{ip6\_mpl.cpp} & 75 {\scriptsize [73--77]} & 97 {\scriptsize [93--97]} & 77 {\scriptsize [76--77]} & $-$20 & $+$2 \\
			\hline
			\multicolumn{6}{l}{\textit{Further files with the largest differences between IF and AFLNet}} \\
			\texttt{mle.cpp} & 384 {\scriptsize [379--424]} & 799 {\scriptsize [789--810]} & 790 {\scriptsize [758--800]} & $-$9 & $+$406 \\
			\texttt{mle\_ftd.cpp} & 356 {\scriptsize [338--401]} & 692 {\scriptsize [645--725]} & 686 {\scriptsize [641--746]} & $-$6 & $+$330 \\
			\texttt{network\_data.cpp} & 18 {\scriptsize [18--18]} & 142 {\scriptsize [117--147]} & 146 {\scriptsize [119--148]} & $+$4 & $+$128 \\
			\texttt{router\_table.cpp} & 68 {\scriptsize [64--95]} & 166 {\scriptsize [133--177]} & 165 {\scriptsize [162--184]} & $-$1 & $+$97 \\
			\texttt{indirect\_sender.cpp} & 1 {\scriptsize [1--1]} & 74 {\scriptsize [49--78]} & 71 {\scriptsize [67--75]} & $-$3 & $+$70 \\
			\hline
		\end{tabular}
	
	\end{table*}

\end{document}